\documentclass[aps,twocolumn,pra,reprint,amsmath,amssymb,floatfix,footinbib,superscriptaddress]{revtex4}

\usepackage{amsmath}        
\usepackage{gensymb}        
\usepackage{upgreek}        
\usepackage{graphics}       
\usepackage{epsfig}         
\usepackage{color}          
\usepackage{bm}             
\usepackage{float}          
\usepackage{ulem}           
\usepackage{blindtext}      

\usepackage{hyperref}        
\hypersetup{colorlinks=true} 

\graphicspath{{./}{./plots/}}

\usepackage{CJK}        
\usepackage{url}
\usepackage{multirow}

\begin{document}

\title{Emergence of quantum spin liquid and spin-flop phase\\ in Kitaev antiferromagnets in a [111] magnetic field}

\author{Shuai Liu}
\affiliation{College of Physics, Nanjing University of Aeronautics and Astronautics, Nanjing, 211106, China}
\affiliation{Key Laboratory of Aerospace Information Materials and Physics (NUAA), MIIT, Nanjing, 211106, China}

\author{Hao Wu}
\affiliation{College of Physics, Nanjing University of Aeronautics and Astronautics, Nanjing, 211106, China}
\affiliation{Key Laboratory of Aerospace Information Materials and Physics (NUAA), MIIT, Nanjing, 211106, China}

\author{Jinbin Li}
\affiliation{College of Physics, Nanjing University of Aeronautics and Astronautics, Nanjing, 211106, China}
\affiliation{Key Laboratory of Aerospace Information Materials and Physics (NUAA), MIIT, Nanjing, 211106, China}

\author{Xiaoqun Wang}
\affiliation{School of Physics and Institute for Advanced Study in Physics, Zhejiang University, Hangzhou 310058, China}

\author{Qiang Luo}
\email[]{qiangluo@nuaa.edu.cn}
\affiliation{College of Physics, Nanjing University of Aeronautics and Astronautics, Nanjing, 211106, China}
\affiliation{Key Laboratory of Aerospace Information Materials and Physics (NUAA), MIIT, Nanjing, 211106, China}

\date{\today}

\begin{abstract}
Kitaev magnets have emerged as pivotal systems for investigating frustrated magnetism, providing a unique platform to explore quantum phases governed by the interplay between bond-dependent anisotropy and external magnetic fields. However, the quantum phase diagrams, particularly near the dominant antiferromagnetic Kitaev regime, remain puzzling despite extensive studies. In this work, we perform unbiased exact diagonalization calculations of the Kitaev-$\Gamma$ model in a [111] magnetic field on a $C_{6}$-symmetric 24-site cluster. By calculating the $\mathbb{Z}_2$ flux density and the topological entanglement entropy, we reveal multiple phase transitions and identify signatures of both scalar and vector chiral orders in the intermediate-field regime between the Kitaev spin liquid and the polarized phase. As the negative $\Gamma$ interaction increases, we discover a proximate quantum spin liquid featured by a three-peak specific heat and a spin-flop phase at a moderate magnetic field. Our findings provide insight into the field-induced intermediate phases in the antiferromagnetic Kitaev model and pave the way for the hunt for emergent phases in real materials.
\end{abstract}


\maketitle

\section{Introduction}

Exchange-anisotropic frustrated magnets stand at the forefront of modern condensed matter physics, serving as a versatile playground for the discovery of exotic quantum phases and a plethora of emergent phenomena \cite{Starykh2015RPP,Cui2023FoP,Rousochatzakis2024RPP,Luo2025CPL}.
In two-dimensional materials, a floating Kosterlitz-Thouless phase is realized in the rare earth magnet TmMgGaO$_4$ with strong Ising anisotropy \cite{Hu2020NC,Li2020NC}, a skyrmion crystal is stabilized in MnSi with a sizable Dzyaloshinskii-Moriya interaction \cite{Muhlbaue2009science, Yi2009PRB}, and a plausible quantum spin liquid (QSL) is identified in Kitaev magnet $\alpha$-RuCl$_3$ under in-plane or out-of-plane magnetic field \cite{Baek2017PRL, Leahy2017PRL, Zheng2017PRL, Banerjee2018npj,Zhou2023NC}.
Among these, the Kitaev interaction, consisting of bond-directional Ising couplings on the honeycomb lattice, stands out because it renders the exact solubility of the Kitaev honeycomb model that can generate the sought-after QSL with fractionalized excitations \cite{Kitaev2006Ann.Phys}.
After a surge of theoretical proposals and/or experimental validation, many Kitaev materials have been proposed, including $5d^5$ iridates \cite{Jackeli2009PRL, Ye2012PRB, Choi2012PRL, Rau2014PRL}, $4d^5$ ruthenates \cite{Sears2015PRB, Kim2015PRB, Johnson2015PRB}, $d^7$ cobaltates \cite{Liu2018PRB,Yao2020PRB,Lin2021NC,Li2022PRX,Jiao2024CM} during the last decade.
Nevertheless, the Kitaev interaction is prevailingly ferromagnetic (FM), and recent efforts have been made to hunt for Kitaev antiferromagnets, prominent examples encompass $\alpha$-RuH$_{3/2}X_{3/2}$ ($X$ = Cl and Br) \cite{Sugita2020PRB}, SmI$_3$ \cite{Xia2025arXiv}, etc.

Theoretically, while the antiferromagnetic (AFM) and FM Kitaev models are unitarily equivalent under local spin rotations at zero field, the AFM model under a magnetic field exhibits significantly richer quantum phase transitions (QPTs) than the FM counterpart, which transitions directly into a polarized phase \cite{Jiang2011PRB}.
However, despite the general recognition of an intervening regime in the AFM Kitaev model in the [111] magnetic field, no broad consensus has yet been reached regarding either the number or the nature of these intermediate phases \cite{Zhu2018PRB,Hickey2019NC, Ronquillo2019PRB,Gohlke2018PRB, Patel2019pnas,Zhang2022NC,Wang2025PRB,Zhu2025NC,Jiang2020PRL,Holdhusen2024PRB,Chen2025PRB}.
While most studies report a single intermediate phase, its proposed nature varies significantly across different methodologies: a gapless $U(1)$ QSL identified through exact diagonalization (ED) \cite{Zhu2018PRB, Hickey2019NC, Ronquillo2019PRB} and density-matrix renormalization group (DMRG) \cite{Zhu2018PRB, Gohlke2018PRB, Patel2019pnas} methods, a gapped chiral QSL with Chern number $C = 4$ from variational approaches \cite{Zhang2022NC}, and a gapless Majorana metal suggested by tensor network studies \cite{Wang2025PRB,Zhu2025NC}. In contrast, combined mean-field and many-body analyses reveal more complex scenarios. Jiang \textit{et al.} identified two distinct disordered phases characterized by $C = 4$ and $C = 1$, respectively \cite{Jiang2020PRL}, while Holdhusen \textit{et al.} reported two symmetry-breaking stripy and chiral phases with either ill-defined or trivial Chern numbers \cite{Holdhusen2024PRB}.
The striking diversity and complexity exhibited by the intermediate phases in the Kitaev model necessitate meticulous inspection through unbiased numerical techniques and sensitive probes of QPTs.

In terms of experiments, the Kitaev QSL (KQSL) has not yet been realized as a genuine ground state of any existing Kitaev material \cite{Matsuda2025arXiv}. Ineluctable exchange couplings, such as the $\Gamma$ interaction originating from the spin-orbit coupling, compete with the Kitaev interaction and typically generate magnetically ordered states \cite{Rau2014PRL}.
The Kitaev-$\Gamma$ model has established itself as a cornerstone model in quantum magnetism, serving as both an important theoretical framework for understanding real materials and a fertile ground for discovering exotic states of matter and unconventional critical behaviors 
\cite{Wang2017PRB,Wang2019PRL,Lee2020NC,Gohlke2020PRR,Luo2024PRB,Zou2025PRB,Rousochatzakis2023KITP,Yang2020PRL,Luo2021PRB,Luo2021PRR,Sorensen2021PRX,Sorensen2024npj}.
When the Kitaev and $\Gamma$ interactions have different signs, the model is highly frustrated and the ground-state phase diagram is rather involved. 
Despite substantial numerical efforts devoted to elucidating QPTs in the FM Kitaev regime, a definitive phase diagram remains unresolved, while intriguing phases including nematic paramagnet \cite{Gohlke2020PRR}, proximate QSL (PQSL) \cite{Wang2019PRL}, and $\Gamma$-type QSL \cite{Luo2021NPJ} have been identified across various parameter ranges (for a recent review, see Ref.~\cite{Rousochatzakis2024RPP}). 
In stark contrast, investigations of QPTs in the AFM Kitaev regime have been remarkably scarce, leaving even the fundamental question of which phase succeeds the KQSL unsettled \cite{Rousochatzakis2023KITP}. This significant knowledge gap, combined with the growing experimental relevance of AFM Kitaev materials, motivates our systematic investigation of the quantum phase diagram in the Kitaev-dominant AFM regime, where we carefully examine the interplay between off-diagonal $\Gamma$ exchange interactions and external magnetic fields.

The paper is structured as follows: 
In Sec.~\ref{SEC:MODEL}, we present the Hamiltonian of the Kitaev-$\Gamma$ model in the out-of-plane magnetic field and introduce two KQSL-related physical quantities, followed by a brief description of the quantum phase diagram.
Section \ref{SEC:RESULT} discusses the features of the QPTs and provides a detailed dynamical and thermal analysis of representative phases. 
Finally, Sec.~\ref{SEC:CONCLUSION} offers a summary of our main results alongside a critical evaluation of the key findings.

\section{Model and Phase Diagram}\label{SEC:MODEL}

We study a spin-$1/2$ AFM Kitaev honeycomb model with a perturbed $\Gamma$ term on a honeycomb lattice in a [111] magnetic field. 
This model is not only experimentally relevant but also theoretically pivotal in exploring exotic phases and emergent excitations.
Its Hamiltonian can be described by 
\begin{align}\label{EQ:Ham}
\mathcal{H} = \sum_{\langle i,j\rangle \parallel \gamma} \big[K S^{\gamma}_{i} S^{\gamma}_{j} + \Gamma \big( S^{\alpha}_{i} S^{\beta}_{j} + S^{\beta}_{i} S^{\alpha}_{j} \big) \big] - \sum_{i} \mathbf{h} \cdot \mathbf{S}_{i},
\end{align}
where $\langle i,j \rangle {\parallel \gamma}$ denotes nearest-neighbor sites connected by a $\gamma$-type bond ($\gamma \in {x,y,z}$), with the interaction type explicitly depending on the bond orientation. $K$ represents the Kitaev interaction, $\Gamma$ denotes the off-diagonal coupling, and the last term corresponds to the magnetic field. 
The magnetic field is applied along the [111] crystallographic direction, and the field strength $\textbf{h}$ is defined via the vector $\textbf{h} = h(1,1,1)/\sqrt{3}$.
In the following analysis, both $K$ and $\Gamma$ are parameterized through an exchange-coupling angle $\phi$, with $K = \mathcal{E}_0 \cos \phi$ and $\Gamma = - \mathcal{E}_0 \sin \phi$ ($\mathcal{E}_0$ is the energy unit). 
The opposing signs of $K$ and $\Gamma$ promote quantum fluctuations, thereby facilitating the emergence of novel quantum states.

Although Hamiltonian Eq.~\eqref{EQ:Ham} has been extensively studied by different many-body computational techniques, the full quantum phase diagram has not yet been settled, especially in the dominant Kitaev region \cite{Rousochatzakis2024RPP}. 
Among these methods, ED is an ideal many-body computational method that provides high accuracy and reliability. It allows a comprehensive study of all relevant physical properties, including thermodynamic and dynamic characteristics.
During the calculation, we chiefly adopted a $C_{6}$-symmetric 24-site cluster with full periodic boundary conditions (PBCs), see Fig.~\ref{FIG-Lattice}(a) in Appendix~\ref{SEC:AppA}. Due to the bond-dependent anisotropy of the Kitaev interactions, a pure lattice $C_6$ rotation is not a symmetry operation of the Kitaev honeycomb model. Even in the isotropic Kitaev limit, the Hamiltonian remains invariant only under a combined operation denoted $C_6 \times C_3^S$, where a $\pi/3$ spatial rotation ($C_6$) is accompanied by a $2\pi/3$ spin rotation about the [111] axis ($C_3^S$).
Therefore, the 24-site cluster we employ is compatible with this combined symmetry structure and thus provides an appropriate finite-size geometry for studying competing phases \cite{Holdhusen2024PRB}.

\begin{figure*}[htb]
\centering
\includegraphics[width=0.90\linewidth, clip]{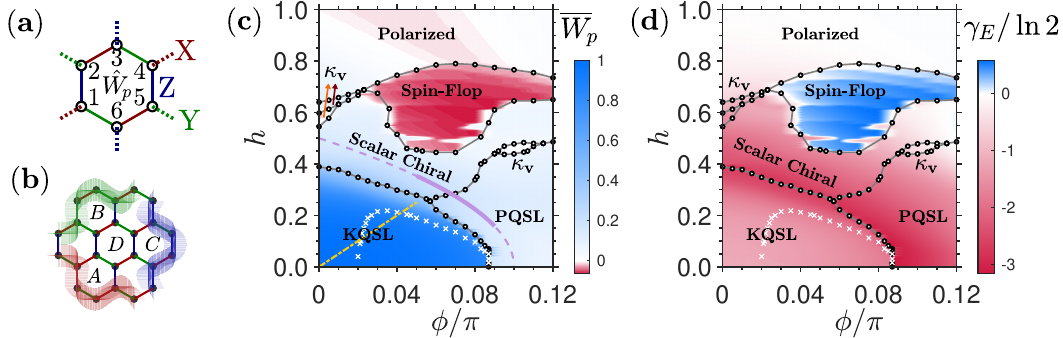}\\
\caption{(a) Illustration of the hexagonal plaquette operator $\hat{W_{p}}$. The nearest-neighbor links are distinguished as \textbf{X} (red), \textbf{Y} (green), and \textbf{Z} (blue) bonds, respectively. (b) The three-region partition scheme employed in calculating topological entanglement entropy using the Kitaev-Preskill method, with the red, green, and blue representing regions $A$, $B$, and $C$, respectively. (c) and (d) present the quantum phase diagrams obtained from the flux density $\overline{W}_{p}$ and TEE $\gamma_E$, respectively. The open circles represent the transition points while the crosses denote a partition within the KQSL. The symbol $\mathbf{\kappa_{v}}$ denotes regions with large vector spin chirality.
The orange and red arrows indicate the regions of the vector chiral phases with ground-state degeneracies of two and three, respectively.
Two special cutting lines, $h = 5 \phi/\pi$ (in yellow) and $h = 0.5\sqrt{1 - \phi/\pi}$ (in purple), are plotted for later use. }
\label{FIG:PhaseDiag}
\end{figure*}

Instead of using energy derivatives and quantum information probes, we will adopt two KQSL-related physical quantities [the flux density and the topological entanglement entropy (TEE)] to characterize QPTs.
On the one hand, the hexagonal plaquette operator, which is schematically illustrated in Fig.~\ref{FIG:PhaseDiag}(a), is defined as
\begin{align}\label{EQ:Wp}
\hat{W_{p}} = 2^{6} S_{1}^{x} S_{2}^{y} S_{3}^{z} S_{4}^{x} S_{5}^{y} S_{6}^{z}.
\end{align}
We compute the expectation value of the plaquette operator for each individual plaquette and define the flux density $\overline{W}_{\!p}$ as its spatial average throughout the entire lattice:
\begin{align}\label{EQ:AverageWp}
\overline{W}_{\!p} = \frac{1}{N_{p}}\sum_{p} \big \langle \hat{W_{p}} \big \rangle,
\end{align}
where $N_{p}$ denotes the total number of plaquettes in the system. The flux density $\overline{W}_{\!p}$ and its derivatives have been frequently used to determine first-order and continuous QPTs \cite{Holdhusen2024PRB, Hickey2019NC, Luo2021NPJ, Feng2023PRB, Wang2025PRB, Lee2020NC}.
On the other hand, we compute the TEE via the Kitaev-Preskill construction \cite{Kitaev2006PRL}. In this recipe, the TEE is computed via a precisely crafted linear combination of von Neumann entropy, given by:
\begin{align}\label{EQ:TEE}
\gamma_{E} = S_{A} + S_{B} + S_{C} - S_{AB} - S_{BC} - S_{AC} + S_{ABC}.
\end{align}
Here, $S_{A}$ represents the von Neumann entropy acquired by tracing out degrees of freedom outside of region $A$, $S_{AB}$ corresponds to the entropy of the union of regions $A$ and $B$, and so on.
To implement this, we partition the cluster into four distinct regions with the same lattice sites, labeled $A$, $B$, $C$, and $D$. The specific choice of these regions is depicted in Fig.~\ref{FIG:PhaseDiag}(b), 
where the red, green, and blue areas correspond to the regions $A$, $B$, and $C$, respectively, while the remaining area defines the region $D$.
Physically, the TEE $\gamma_E$ remains constant within a given phase but exhibits variations across distinct phases. Consequently, the discontinuity in TEE serves as a sensitive indicator of QPTs \cite{Ronquillo2019PRB, Holdhusen2024PRB, Feng2023PRB}.

Figure~\ref{FIG:PhaseDiag}(c) and \ref{FIG:PhaseDiag}(d) present the quantum phase diagrams characterized by the flux density $\overline{W}_{p}$ and the TEE $\gamma_E$, respectively. The concurrence of singularities in both quantities unambiguously identifies consistent phase boundaries. 
The quantum phase diagrams in the $(\phi/\pi, h)$ parameter space predominantly consist of five distinct phases, which include the KQSL, a proximate QSL featured by a three-peak specific heat, a scalar chiral order and a spin-flop phase at moderate magnetic field, and a polarized phase at high magnetic field. There are also three narrow regions of vector chiral order near $\phi/\pi$ = 0.00 and 0.10. 
For comparison, the density plots of the scalar and vector order parameters are presented in the Appendix ~\ref{SEC:AppA}.
In what follows, we will systematically investigate the characteristic properties of each individual phase, with particular focus on their spin and thermal dynamics.

\section{Results and Discussion}\label{SEC:RESULT}

\subsection{Competing degenerate states in KQSL}

The Kitaev honeycomb model provides a rare, exactly solvable platform where the spins fractionalize into Majorana fermions.
At zero field, this leads to a gapless QSL with an emergent $\mathbb{Z}_2$ gauge structure \cite{Kitaev2006Ann.Phys}.
These Majorana quasiparticles are classified into two kinds, itinerant Majorana fermions and flux excitations (so-called visons), both of which play key roles at different energy and temperature scales.
In the $\mathbb{Z}_2$ QSL, the ground state lies in the vortex-free sector in which $\langle \hat{W}_p\rangle = 1$ for arbitrary hexagonal plaquette, yielding flux density $\overline{W}_p = 1$.
The TEE $\gamma_E = -\ln 2$, as the $\mathbb{Z}_2$ topological order has a quantum dimension $\mathcal{D} = 2$.
Beyond the ground state, hallmark signatures of the two kinds of Majorana fermions manifest in spin dynamics and thermal measurements. The itinerant Majorana fermions and visons contribute, respectively, to the broad high-energy continuum and the low-energy coherent peak in the dynamical spin structure factor (DSSF) \cite{Knolle2014PRL,Knolle2015PRB}.
They also account for a double-peak structure in specific heat and a nearly half-plateau thermal entropy in between \cite{Nasu2015PRB}.

\begin{figure}[htbp]
\centering
\includegraphics[width=0.95\columnwidth, clip]{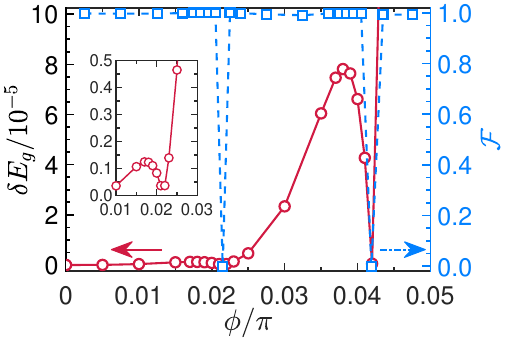}\\
\caption{The energy-level splitting $\delta E_g$ (left axis, red circles) and the ground-state fidelity $\mathcal{F}(\phi, h)$ (right axis, blue squares) as a function of $\phi$ along the parameter path $h = 5\phi/\pi$. The inset magnifies $\delta E_g$ in the region where $0.01 \leq \phi/\pi \leq 0.03$.}
\label{FIG:FidelityvsGap}
\end{figure}

However, the exact solvability of the Kitaev honeycomb model can be 
ruined immediately by perturbations such as magnetic field or off-diagonal exchanges. Numerical calculations are therefore necessary to quantify the extent of the KQSL region.
As shown in Figs.~\ref{FIG:PhaseDiag}(c) and \ref{FIG:PhaseDiag}(d), the phase boundary of the KQSL in the ($\phi/\pi$, $h$) parameter plane
is consistently delineated by using flux density $\overline{W}_p$ and TEE $\gamma_E$.
Specifically, the KQSL ends at $h_{t,l} \approx 0.385$ and $\phi_{t}/\pi \approx 0.087$, respectively, in the absence of the $\Gamma$ term and in the zero-field Kitaev-$\Gamma$ model.
These transition point estimates show excellent consistency with the previous ED study \cite{Hickey2019NC,Rousochatzakis2023KITP}.

\begin{figure}[!ht]
\centering
\includegraphics[width=0.95\columnwidth, clip]{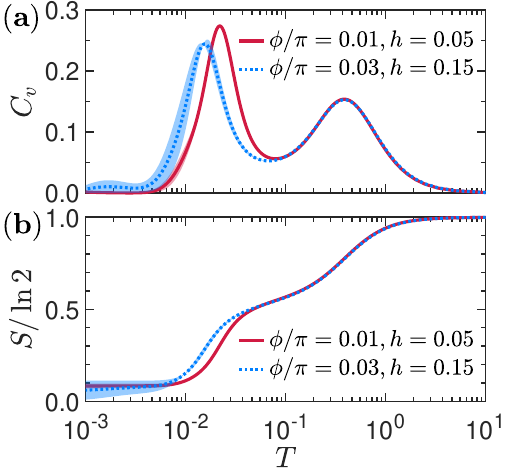}\\
\caption{(a) and (b) show the specific heat $C_v$ and the thermal entropy $S$ at two characteristic parameter points, ($\phi / \pi = 0.01$, $h = 0.05$) and ($\phi / \pi = 0.03$, $h = 0.15$), along the path line $h =5 \phi/\pi$. The shaded regions mark the estimated errors in the low-temperature regime.}
\label{FIG:KSLCvEntropy}
\end{figure}

Although KQSL is characterized by a four-fold degeneracy of the ground state on a torus \cite{Kurita2015PRB}, the degenerate manifold can be lifted in a finite 24-site cluster. In accordance with Rousochatzakis \cite{Rousochatzakis2023KITP}, we also observe that the ground-state degeneracy is three in the pure Kitaev limit and is reduced to two as the $\Gamma$ term is added.
Interestingly, there is still competition in the presence of the magnetic field, resulting in a tiny energy-level splitting between the remaining two degenerate states.
To illustrate the evolution of the ground state with the lowest energy, we have calculated the ground-state fidelity $\mathcal{F}(\phi, h) = | \langle \psi_{g}(\phi', h') | \psi_{g}(\phi, h) \rangle |$ in which $(\phi', h')$ close to $(\phi, h)$, and the energy-level splitting $\delta E_g = E_1 - E_0$ with $E_{0,1}$ being the energy of the two lowest energy levels.
Figure~\ref{FIG:FidelityvsGap} shows the splitting $\delta E_g$ (left axis, in red) and the fidelity $\mathcal{F}$ (right axis, in blue) along the parameter line $h = 5 \phi/\pi$ with $0\leq \phi/\pi \leq 0.05$.
It is clear that $\delta E_g$ has two appreciable drops down to zero when $\phi_{s,1}/\pi \approx$ 0.022 and $\phi_{s,2}/\pi \approx$ 0.042, signifying an accidental balance due to the interplay of competing interactions and magnetic field.
Interestingly, $\mathcal{F}$ approaches zero at $\phi_{s,1}$ and $\phi_{s,2}$, indicating that there is a swap of the degenerate ground state thereof.

Although we are not aware of any physical implication of competition of the degenerate ground state at present, this effect breaks the KQSL into two separated regions, as marked by white crosses ($\times$) in Figs.~\ref{FIG:PhaseDiag}(c) and \ref{FIG:PhaseDiag}(d).
Here, we resort to thermodynamic quantities, such as specific heat and thermal entropy, to testify to the similarity of the two regions.
The thermodynamic properties were computed using the canonical thermal pure quantum (TPQ) state method \cite{Sugiura2013PRL}, as implemented in the open-source software package $\mathcal{H} \phi$ \cite{Kawamura2017CPC, Ido2024CPC}. In the TPQ method, the $k$-th TPQ state is generated as $|\phi_{\text{TPQ}} (\beta_k) \rangle = \left[ U(\Delta \tau) \right]^k | \phi_{\text{rand}} \rangle$ where the inverse temperature $\beta_k = k \Delta \tau$ and the initial state $ | \phi_{\text{rand}} \rangle $ is a random vector. In addition, $ U(\Delta \tau) = e^{-\frac{\Delta \tau}{2} \mathcal{H}} \approx \sum_{n=0}^{n_{\max}} \frac{1}{n!} \left( -\frac{\Delta \tau}{2} \mathcal{H} \right)^n $. Here, $n_{\max}$ denotes the maximal order of the Taylor expansion and is set to 20, and $\Delta \tau$ represents the increment of the imaginary-time evolution and is set to 0.1. 
To ensure statistical reliability, stochastic sampling was carried out over three independent runs, and the final results were obtained by averaging the outcomes of these executions.

To commence with, the essential step is to determine the free energy $F$ of the system, which serves as the basis for all subsequent thermodynamic computations. It is known that $F = -\beta^{-1} \ln {\mathcal{Z}}$, where $\mathcal{Z}$ denotes the partition function of the system and $\beta = 1/k_{\text{B}}T$ (with the Boltzmann constant $k_{\text{B}}$ set to 1 in this paper). Based on this, the specific heat and the thermal entropy are calculated as
\begin{subequations}\label{EQ:CvS}
\begin{align}
C_{v} = \frac{1}{N} \Big ( \frac{\partial U}{\partial T}  \Big )_{v}, \label{EQ:SpecificHeat} \\
S = \frac{\beta}{N} \left (U - F\right), \label{EQ:ThermalENtropy}
\end{align}
\end{subequations}
with the internal energy $U = -{\partial \ln{\mathcal{Z}}}/{\partial \beta}$ \cite{Luo2021PRR}.
We have chosen a representative point in each region with $\phi/\pi$ = 0.01 and 0.03 along the line $h = 5 \phi/\pi$.
As can be seen in Fig.~\ref{FIG:KSLCvEntropy}, the specific heat displays two pronounced peaks at two well-separated temperatures due to two kinds of fractionalized excitations, and the thermal entropy is released in two steps with a quasi-half-plateau in the middle. These observed features are consistent with the expected signatures of a KQSL.

Before closing, we briefly note that a spurious lattice nematicity was identified in low-field KQSL. This behavior stems from ground-state degeneracy and is highly sensitive to numerical precision. More details are provided in Appendix~\ref{SEC:AppB}.

\begin{figure}[!ht]
\centering
\includegraphics[width=0.95\columnwidth, clip]{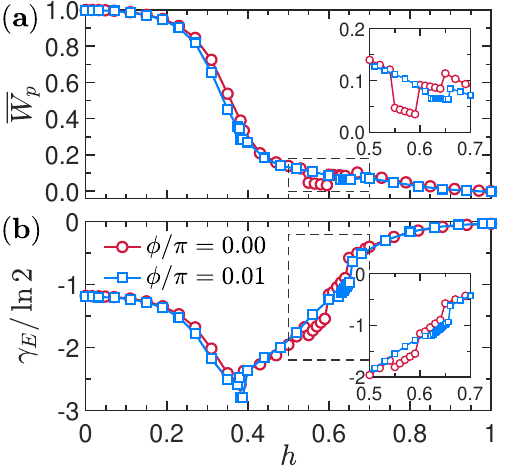}\\
 \caption{(a) The flux density $\overline{W}_{p}$ as a function of $h$ along parameter lines $\phi/\pi = 0.00$ (red circles) and 0.01 (blue squares). The insets show a magnified view in the range of $0.5 \leq h \leq 0.7$. (b) The same as (a) but for the TEE $\gamma_E$.}
\label{FIG:Wp&TEE_HcTuned}
\end{figure}

\subsection{Field-induced scalar and vector chiral phases}

Having established the phase boundary of the KQSL, we now consider the emergent phases in the quantum phase diagram.
Aligning with the enormous efforts to identify the field-induced intermediate region in the Kitaev model in the out-of-plane magnetic field, we made an attempt to provide insight into the number and nature of the intermediate phases.
As can be seen in Fig.~\ref{FIG:Wp&TEE_HcTuned}, flux density $\overline{W}_p$ and TEE $\gamma_E$ undergo several abrupt changes as the magnetic field $h$ increases to 1, at which the underlying ground state becomes polarized.
Taking $\phi/\pi = 0$ as an example, $\overline{W}_p$ decreases from unity and has its first jump at $h_{t}^{l} \approx 0.385$. 
Subsequently, it experiences two jumps at $h_{t}^{m_1} \approx 0.540$ and $h_{t}^{m_2} \approx 0.595$, before entering the polarized phase at $h_{t}^{h} \approx 0.640$.
Similarly, these abrupt changes are also observed in the curve of $\gamma_E$, verifying the existence of multiple QPTs.
Beyond the pure Kitaev limit, key features are observed to be still reserved at a slightly different angle $\phi/\pi = 0.005$ (not shown), ruling out the possibility of numerical instability. Hence, the intermediate regions contain three parts, in which one is quite wide while others are relatively narrow. 
We also find that the intermediate regions have a large number of low-energy states.
However, as $\phi/\pi$ increases to 0.01, one of the narrow regions disappears, see the inset.

We proceed to reveal the nature of these intermediate phases.
Magnetically ordered states are known to be distinguished by their patterns in the static spin structure factor (SSSF), which is given by $\mathbb{S}_N({\bf{q}}) = \sum_{\alpha\beta}\delta_{\alpha\beta}\mathbb{S}_N^{\alpha\beta}({\bf{q}})$, where
\begin{equation}\label{EQ:SSF}
\mathbb{S}_N^{\alpha\beta}({\bf{q}})=\frac{1}{N}\sum_{ij} \langle{S^{\alpha}_i {S^{\beta}_j}}\rangle e^{\imath{\bf{q}}\cdot{({\bm{R}}_i-{\bm{R}}_j)}}.
\end{equation}
We find that the SSSFs in these intermediate phases are rather diffusive without pronounced peaks, eliminating the likelihood of long-range orders.
In order to look for possible symmetry-breaking signatures, we measured two kinds of chiralities: scalar spin chirality and vector spin chirality.

The scalar spin chirality, which breaks time-reversal symmetry, is characterized by a scalar chiral order parameter denoted as 
\begin{align}\label{EQ:S_Chiral}
\chi_{i,j,k} = 2^3 \left \langle \mathbf{S}_{i} \cdot \left ( \mathbf{S}_{j} \times \mathbf{S}_{k} \right ) \right \rangle = 2^3 \sum_{\{\alpha\beta\gamma\}} \eta^{\alpha\beta\gamma} \langle S_{i}^{\alpha} S_{j}^{\beta} S_{k}^{\gamma} \rangle,
\end{align}
where the sites $\left(i, j, k \right)$ of the same sublattice form an equilateral triangle in the anticlockwise direction. $\eta^{\alpha\beta\gamma} = \pm1$, depending on the permutation order.  
As the magnetic field is applied, the chirality is usually nonzero because of the time-reversal symmetry breaking. To deal with distraction, we substitute the third-order moment in Eq.~\eqref{EQ:S_Chiral} with a cumulant:
\begin{align}\label{EQ:ThirdOrder}
\kappa_{3}\big( S_{i}^{\alpha} S_{j}^{\beta} S_{k}^{\gamma}\big) = 
&\big\langle S_{i}^{\alpha} S_{j}^{\beta} S_{k}^{\gamma} \big\rangle 
- \big\langle S_{i}^{\alpha} S_{j}^{\beta} \big\rangle \big\langle S_{k}^{\gamma} \big\rangle \nonumber \\
&- \big\langle S_{i}^{\alpha} S_{k}^{\gamma} \big\rangle \big\langle S_{j}^{\beta} \big\rangle 
 - \big\langle S_{j}^{\beta} S_{k}^{\gamma} \big\rangle \big\langle S_{i}^{\alpha} \big\rangle \nonumber \\
&+ 2\big\langle S_{i}^{\alpha} \big\rangle \big\langle S_{j}^{\beta} \big\rangle \big\langle S_{k}^{\gamma} \big\rangle.
\end{align}
In doing so, we systematically eliminate contributions from lower-order statistics (single and pairwise correlations), thereby rigorously extracting the nontrivial correlation effects solely arising from three-body interactions. This approach not only guarantees that the scalar spin chirality signal originates exclusively from intrinsic three-body quantum correlations but also renders it a sensitive probe for characterizing time-reversal symmetry breaking.
Ultimately, the scalar chiral order parameter is defined as

\begin{align}\label{EQ:ScaChiNew}
\chi_{s} = \frac{2}{N} \sum\chi_{i,j,k} ,
\end{align}
where the summation is taken over all eligible triangles within the hexagonal plaquettes.

\begin{figure}[!ht]
\centering
\includegraphics[width=0.95\columnwidth, clip]{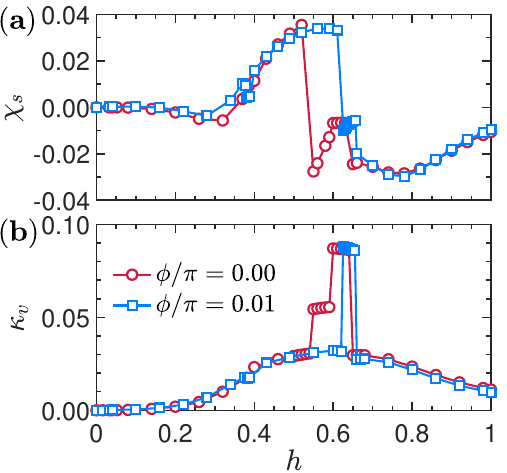}\\
\caption{(a) The scalar chiral order parameter $\chi_s$ as a function of $h$ along parameter lines $\phi/\pi = 0.00$ (red circles) and 0.01 (blue squares). (b) The same as (a) but for the vector chiral order parameter $\mathbf{\kappa_v}$.}
\label{FIG-ChiralOPs}
\end{figure}

The vector spin chirality that corresponds to the 
breaking of discrete symmetry is given by
\begin{align}\label{EQ:V_Chiral}
&\kappa_{i,j}^{\gamma} = \left\langle S_i^{\alpha}S_j^{\beta} -  S_i^{\beta}S_j^{\alpha} \right\rangle, 
\end{align}
where the sites $\left(i, j\right)$ are of next-nearest neighbor and their bond type $\gamma$ is the same as the corresponding orthogonal bonds.
As the discrete symmetry breaks spontaneously, the ground state $\vert\psi\rangle$ should have a degeneracy $g$ and the chirality is then averaged with respect to the degenerate manifold.
Hence, the vector chiral order parameter is defined as
\begin{align}\label{EQ:VecChiNew}
\mathbf{\kappa_{v}} = \frac{1}{3gN} \sum_{\langle\langle i,j\rangle\rangle} \sum_{\psi} \left(\kappa_{i,j}^{\gamma}\right)_{\psi}.
\end{align}

\begin{figure}[!ht]
\centering
\includegraphics[width=0.95\columnwidth, clip]{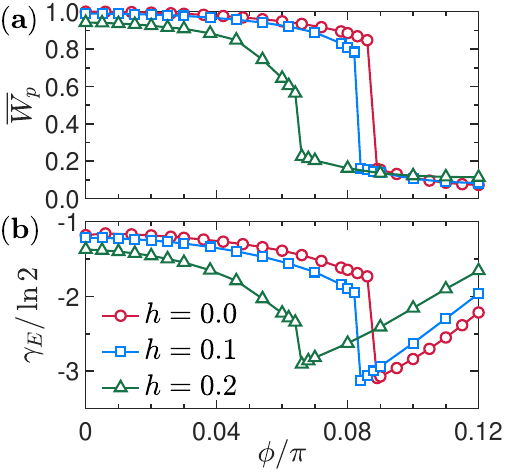}\\
\caption{(a) The flux density $\overline{W}_{p}$ as a function of $\phi$ along parameter lines $h = 0.0$ (red circles), 0.1 (blue squares), and 0.2 (green triangles). (b) The same as (a) but for the TEE $\gamma_E$.}
\label{FIG:Wp&TEE_PhiTuned}
\end{figure}

The scalar ($\chi_s$) and vector ($\mathbf{\kappa_v}$) chiral order parameters are shown in Figs.~\ref{FIG-ChiralOPs}(a) and \ref{FIG-ChiralOPs}(b), respectively.
In the case of $\phi/\pi = 0.00$, $\chi_s$ remains negligible in the KQSL. It exhibits a sharp increase within the field range $h \in (0.385, 0.540)$, attaining a peak value of $\sim 0.04$. Subsequently, $\chi_s$ undergoes an abrupt drop accompanied by a sign reversal, eventually decaying to zero in the polarized phase.
The phase adjacent to the KQSL, characterized by a maximal scalar chiral order parameter, is designated as the scalar chiral phase.
In contrast, $\kappa_v$ displays a dome-like behavior as $h$ varies, with the most striking feature being two quasi-plateaus of $~0.055$ and $~0.095$ in the field ranges $h \in (0.540, 0.595)$ and $h \in (0.595, 0.640)$, respectively.
The ground state degeneracies $g$ of the two intervals are two and three at low and high fields, respectively.
Despite these differences, the enhanced order parameter observed in intermediate-field regimes indicates the emergence of vector chiral correlations within a narrow field range below saturation. Consequently, we identify the regions that exhibit the maximal chiral order parameter as vector chiral phases.
As $\phi/\pi$ increases, the domain of the scalar chiral phase expands, whereas those of the vector chiral phases shrink. Notably, the twofold-degenerate vector chiral phase completely vanishes at $\phi/\pi = 0.01$. Our findings highlight the need for careful verification of vector chiral phases in future large-scale many-body calculations that surpass the ED limitations.
It would be worthwhile for such future studies to investigate whether such a phase constitutes a chiral QSL or a symmetry-breaking chiral spin solid.

\begin{figure}[!ht]
\centering
\includegraphics[width=0.95\columnwidth, clip]{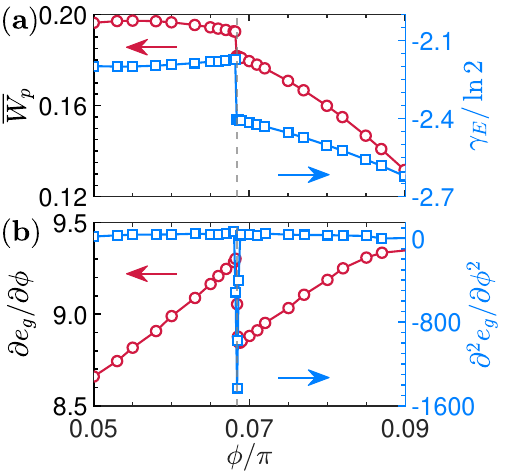}\\
\caption{(a) The flux density $\overline{W}_{p}$ (left axis, red circles) and the TEE $\gamma_E$ (right axis, blue squares) as a function of $\phi$ along the parabolic parameter path $h = 0.5\sqrt{1 - \phi/\pi}$. (b) The same as (a) but for the first-order (${\partial e_{g}}/{\partial \phi}$, left axis) and second-order (${\partial^2 e_{g}}/{\partial \phi ^2}$, right axis) energy derivatives.}
\label{FIG-Parabla}
\end{figure}

Finally, we comment on why vector chiral phases may have been overlooked in previous ED studies of 24-site clusters.
In Ref.~\cite{Hickey2019NC}, although the same $C_6$-symmetric 24-site cluster is employed, one of the main cases studied involves a magnetic field tilted away from the $\textbf{c}$ axis ($[111]$ direction) by an angle $\vartheta = 7.5^\circ$. As shown in Appendix ~\ref{SEC:AppC}, the vector chiral phases indeed vanish at this tilt angle.
In Ref.~\cite{Zhu2018PRB}, a different 24-site torus ($2 \times 4 \times 3$) is used. Nevertheless, as can be seen from Appendix~\ref{SEC:AppD}, a finite—though reduced—chiral region can still be resolved when the magnetic field resolution is increased.
These findings highlight that subtle aspects of the numerical setup, including field direction, lattice topology, and computational resolution, are critical for reliably resolving exotic magnetic phases.

\begin{figure*}[htb]
\centering
\includegraphics[width=0.90\linewidth, clip]{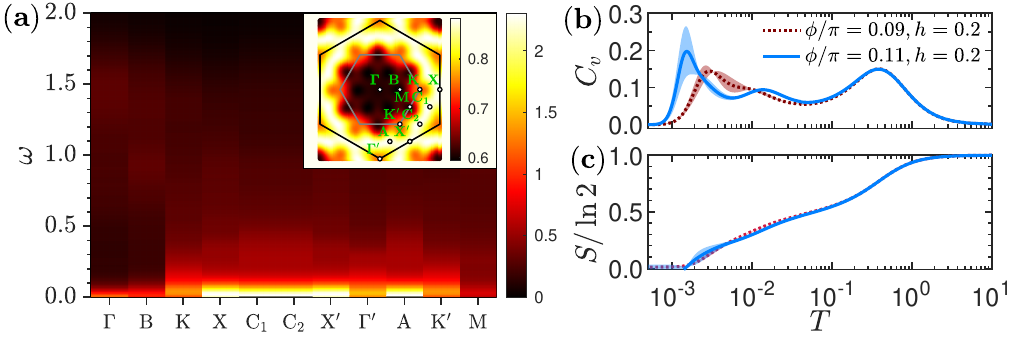}\\
\caption{(a) DSSF calculated along high-symmetry momentum lines at $(\phi/\pi, h) = (0.11, 0.2)$. Inset: The corresponding SSSF with labels of symmetry points used in the main panel. (b) and (c) show the specific heat $C_v$ and the thermal entropy $S$ at two characteristic parameter points, ($\phi / \pi = 0.09$, $h = 0.2$) and ($\phi / \pi = 0.11$, $h = 0.2$). The shaded regions mark the estimated errors in the low-temperature regime. The observed negative entropy values at ultralow temperatures are attributed to systematic errors resulting from exceeding the precision limit of the measurement.}
\label{FIG-HighGamma}
\end{figure*}

\subsection{Proximate QSL: high TEE and three-peak specific heat}

We now move on to further consider the role of the negative $\Gamma$ interaction. As $|\Gamma|$ increases, the KQSL is destroyed and taken over by a new phase. Similarly, we leverage the flux density $\overline{W}_{p}$ and the TEE $\gamma_E$ to pinpoint the phase boundary.
As can be seen in Fig.~\ref{FIG:Wp&TEE_PhiTuned}, the KQSL gives way to a new phase through a first-order QPT at $h$ = 0.0 (red circles), 0.1 (blue squares), and 0.2 (green triangles).
Compared to the transition between the KQSL and the scalar chiral phase revealed in Fig.~\ref{FIG:Wp&TEE_HcTuned}, the jumps in $\overline{W}_{p}$ and $\gamma_E$ here are much sharper, especially in the low field case.
Despite behavioral differences, the new phase exhibits certain similarities to the scalar chiral phase, as evidenced by the flux density $\overline{W}_{p}$ decreasing below 0.1 and the rescaled TEE $|\gamma_E|/\ln2$ decreasing from an initial value of approximately three.
This raises the curious question of whether a QPT exists between the scalar chiral phase and the new phase. 
To address this issue, we designed a parabolic parameter path $h = 0.5\sqrt{1 - \phi/\pi}$ and performed comprehensive calculations of the flux density $\overline{W}_{p}$ and the TEE $\gamma_E$ along this trajectory; see Fig.~\ref{FIG-Parabla}(a). As the exchange-coupling angle $\phi$ is varied, discontinuous jumps in both quantities are observed, signaling the occurrence of a first-order QPT.
To further corroborate the nature of the transition, we computed the first and second derivatives of the ground-state energy ($\partial  e_g/\partial\phi$ and $\partial^2  e_g/\partial\phi^2$), which exhibit a discontinuity and a divergent peak, respectively, consistent with a first-order QPT [see Fig.~\ref{FIG-Parabla}(b)].
Thus, all the calculations demonstrate that the new phase is fundamentally distinct from the scalar chiral phase.

To reveal the nature of this newly identified phase, we employ complementary diagnostics to the SSSF, which exhibits a diffusive pattern in reciprocal space, indicating the absence of conventional long-range order.
The DSSF is particularly noteworthy, as it is not only capable of probing fractionalized excitations but also provides direct signatures measurable in spectroscopic experiments \cite{Laurell2020npj}.
The DSSF is calculated from the Fourier transform in time and space of the dynamical spin correlation function
$\mathbb{S}_{i,j}^{\gamma \gamma}\left ( t \right ) = \left \langle S_{i}^{\gamma}\left ( t \right ) S_{j}^{\gamma}\left ( 0 \right ) \right \rangle$
where $S_{i}^{\gamma}(t)$ is the $\gamma$-th component of the spin operator in the Heisenberg representation at time $t$ on site $i$. It is formulated as
\begin{align}\label{EQ:DSF}
\mathbb{S}_{\mathbf{q}}^{\gamma\gamma}\left ( \mathbf{\omega} \right ) = \frac{1}{N} \sum_{i,j} e^{i \mathbf{q}\left ( \mathbf{r}_{i} - \mathbf{r}_{j} \right )} \int_{-\infty }^{\infty} dt e^{-i \mathbf{\omega} t} \mathbb{S}_{i,j}^{\gamma \gamma}\left ( t \right ).
\end{align}
Considering the zero-temperature case, the average is taken in the ground state and the Lehmann representation of the DSSF is
\begin{align}\label{EQ:Lehmann}
\mathbb{S}_{\mathbf{q}}^{\gamma \gamma }(\omega) = \sum_{\upsilon}
\left| \langle \upsilon | S_{\mathbf{q}}^{\gamma} | 0  \rangle \right|^2 \delta \big( \mathbf{\omega} - ( E_{\upsilon} - E_{0}) \big),
\end{align}
where $E_{\upsilon}$ is the energy corresponding to an eigenstate $\vert\upsilon\rangle$ \cite{Knolle2015PRB}.
We compute the DSSF using its continued fraction representation. This is achieved by preparing an initial state, which is then processed through the Lanczos algorithm. The resulting tridiagonal matrix implicitly contains the coefficients of the continued fraction expansion \cite{Gagliano1987PRL}. In our implementation, we set the energy resolution $\delta \omega = 0.01$, employ a Gaussian broadening parameter $\eta = 0.05$, and perform $1000$ Lanczos iterations to ensure numerical convergence of the DSSF. Furthermore, from now on, we will focus on the sum $\mathbb{S}\left({\mathbf{q}}, \mathbf{\omega} \right) = \sum_{\gamma } \mathbb{S}_{\mathbf{q}}^{\gamma \gamma }\left ( \mathbf{\omega} \right )$.

To be specific, we will examine certain parameter points along the line of $h = 0.2$. In this new phase, the ground state is unique and the energy spectrum is very dense with a tiny excitation gap. Taking $\phi/\pi = 0.11$ which lies well within the bulk of the phase as an example, we find an excitation gap of $\Delta \approx 0.008$.
The SSSF and the DSSF at the same parameter point are shown in Fig.~\ref{FIG-HighGamma}(a). For the SSSF shown in the inset, it is observed that the high intensity is diffusely distributed over a wide momentum range, with the nominal peak located halfway between the vertices of the first and second Brillouin zones. Meanwhile, the DSSF exhibits broad and continuous spectral weight characteristics in the low-frequency range. 
The slight spectral weight at high-frequency regime around $\omega \approx 1.5$ may be attributed to either fractionalized excitations or finite-resolution effects.
Together, these phenomena indicate the presence of a likely QSL in the ground state.

To detect possible unusual excitations, we compute the specific heat $C_v$ and thermal entropy $S$ at the parameter points $\phi/\pi = 0.09$ and $\phi/\pi = 0.11$, which are close to the phase boundary and deep into the new phase, respectively.
As shown in Fig.~\ref{FIG-HighGamma}(b), the specific heat curve at $\phi/\pi = 0.09$ (in red) shows not only low-temperature and high-temperature peaks but also a smeared intermediate peak. As $\phi$ increases, say to $\phi/\pi = 0.11$, the high-temperature peak remains stable while the low-temperature peak shifts to a lower temperature and becomes more pronounced. Notably, the intermediate peak is also enhanced and evolves into a distinctly observable feature.
In parallel, the thermal entropy, depicted in Fig.~\ref{FIG-HighGamma}(c), undergoes a three-step release in accordance with the specific heat having a three-peak structure.
We note that such a three-peak specific heat is scarce and the multiple peaks may be a reminiscence of fractionalized excitations.
Compared with neighboring KQSL, we conclude that the new phase is a proximate QSL characterized by high TEE and three-peak specific heat.

\subsection{The spin-flop phase: double peaks and logarithmic temperature scaling in specific heat}

The discovery and characterization of magnetic phase transitions in low-dimensional magnetic systems are pivotal to advancing our fundamental understanding of magnetism. Among these, the spin-flop transition represents a cornerstone in the physics of antiferromagnetism, offering profound insights into the interplay between exchange interactions, anisotropy, and external fields.
The spin-flop transition is a magnetic phase transition in which the spins of an easy-axis antiferromagnet undergo a sudden reorientation under an applied magnetic field, giving rise to a transverse magnetization.
Such a transition has been probed in honeycomb-lattice materials including Ni$_2$Mo$_{3}$O$_{8}$ \cite{Yu2024PRB}, CaCo$_{2}$TeO$_{6}$ \cite{Huang2025PRB}, Na$_{3}$Ni$_{2}$BiO$_{6}$ \cite{Shangguan2021NP}, and FeP$_3$SiO$_{11}$ \cite{Khatua2024PRB}.

\begin{figure}[!ht]
\centering
\includegraphics[width=0.95\columnwidth, clip] {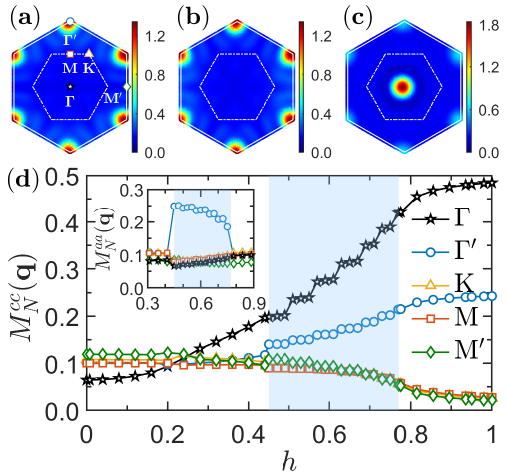}\\
\caption{(a), (b), and (c) respectively display the SSSFs for spin components along three crystallographic directions $\textbf{a} [11\bar2]$, $\textbf{b} [\bar110]$, and $\textbf{c} [111]$ at $(\phi/\pi = 0.06, h = 0.6)$.
Panel (a) specifically marks the selected momentum points for magnetic order parameter calculations.
(d) Evolution of magnetic order parameters 
$M_N^{cc}(\boldsymbol{\Gamma})$ (black star), $M_N^{cc}(\boldsymbol{\Gamma}')$ (blue circle), $M_N^{cc}(\textbf{K})$ (yellow triangle), 
$M_N^{cc}(\textbf{M})$ (orange square), and $M_N^{cc}(\textbf{M}')$ (green diamond) along the \textbf{c}-axis at $\phi / \pi = 0.06$. The inset shows the corresponding results for the \textbf{a}-axis direction.}
\label{FIG-spinflop}
\end{figure}

However, different from the conventional wisdom that the spin-flop phase is induced from a collinear AFM phase, we find that it can alternatively be stabilized at a moderate magnetic field even in the absence of a low-field magnetically ordered state.
To demonstrate it, we present QPTs along the line of $\phi/\pi = 0.06$ in which the zero-field ground state is KQSL.
Figs.~\ref{FIG-spinflop}(a)-\ref{FIG-spinflop}(c) display the SSSFs at $h = 0.6$ for spin components along $\textbf{a} [11\bar2]$, $\textbf{b} [\bar110]$, and $\textbf{c} [111]$, respectively.
The out-of-plane SSSFs exhibits a dominant peak at the $\Gamma$ point, whereas the in-plane SSSFs have the same intensity and peak at the $\Gamma'$ point. These magnetization behaviors demonstrate that the spins of the two sublattices are antiparallel when projected onto the honeycomb plane, while exhibiting equal canting angles toward the $\textbf{c}$-axis.
Next, we calculate the magnetic order parameter $M_N^{\gamma\gamma}(\textbf{q}) = \sqrt{\mathbb{S}_N^{\gamma\gamma}({\bf{q}})/N}$ at different high-symmetry points, where $\mathbb{S}_N^{\gamma\gamma}({\bf{q}})$ denotes the SSSF defined in Eq.~\eqref{EQ:SSF}. 
As shown in Fig.~\ref{FIG-spinflop}(d), $M_N^{\gamma\gamma}(\textbf{q})$ with $\textbf{q}$ = $\textbf{K}$ (yellow triangle), $\textbf{M}$ (orange square), and $\textbf{M}'$ (green diamond) are smaller than 0.1 and are considerably reduced as $h$ increases, excluding the presence of 120$^{\circ}$ order, zigzag order, and stripy order. In contrast, $M_N^{\gamma\gamma}(\boldsymbol{\Gamma})$ (black star) and $M_N^{\gamma\gamma}(\boldsymbol{\Gamma}')$ (blue circle) increase simultaneously with the increase of $h$. In particular, the 
uniform magnetization $M_N^{\gamma\gamma}(\boldsymbol{\Gamma})$ increases in steps in the interval of $h\in (0.445, 0.775)$, a manifestation of the reorientation of the underlying spins against the magnetic field. In comparison, the in-plane  antiferromagnetic order parameters $M_N^{aa}(\boldsymbol{\Gamma}')$ and $M_N^{bb}(\boldsymbol{\Gamma}')$ in this interval are sizable (see the inset), albeit having a gradual reduction towards the saturated magnetic field.
The onset of in-plane
staggered magnetization associated with the $\boldsymbol{\Gamma}'$ point, together with the rising out-of-plane 
uniform magnetization associated with the $\boldsymbol{\Gamma}$ point, convincingly verifies the formation of the spin-flop phase.

\begin{figure}[!ht]
\centering
\includegraphics[width=0.95\columnwidth, clip] {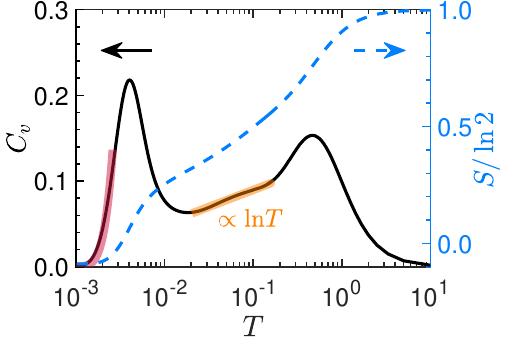}\\
\caption{The specific heat $C_v$ (left axis, solid line) and the thermal entropy $S$ (right axis, dotted line) at the parameter points ($\phi / \pi = 0.06$, $h = 0.6$). In the specific heat, the low-temperature asymptotic behavior ($e^{-0.012937/T}/T^{1/2}$, red belt) and a logarithmic scaling behavior ($0.017\ln T + 0.129$, orange belt) in the middle temperature region are plotted.}
\label{FIG-spinflopCVandS}
\end{figure}

The coexistence of transverse and longitudinal magnetizations in the spin-flop phase may contribute to the creation of different magnetic excitations. To measure the possible multiple excitations, we here calculate the specific heat $C_v$ and the thermal entropy $S$ at the representative point $(\phi/\pi, h) = (0.06, 0.6)$ in the spin-flop phase.
As can be seen in Fig.~\ref{FIG-spinflopCVandS}, the specific heat exhibits two well-separated peaks at different temperature scales, in which the low temperature is two orders smaller than the high one. Although the low temperature is small, posing experimental challenges, the peak is nevertheless robust with a high intensity. 
For the 24-site cluster, we find that the 
low-temperature specific heat behaves 
approximately as $C_v \simeq e^{-\Delta_s/T}/T^{\alpha_s}$ with $\Delta_s \approx 0.012937$ being the excitation gap and $\alpha_s = 1/2$.
In addition, the specific heat exhibits a logarithmic temperature dependence within the intermediate regime bounded by the two characteristic peaks.
This logarithmic scaling behavior is a hallmark signature of the well-known superfluid phase.
Indeed, the correspondence between the spin-flop phase and the superfluid phase has been demonstrated in an analogous spin-orbit coupled model \cite{Luo2022PRB}. The distinction between the two spin-flop phases manifests in the relative alignment of their in-plane spin components: parallel (ferromagnetic spin-flop phase) versus antiparallel (antiferromagnetic spin-flop phase) configurations \cite{Li2025PRB}.

Our results incidentally clarify that the appearance of double peaks in the specific heat cannot serve as definitive evidence for QSLs, contrary to some previous claims in the literature \cite{Xu2020PRL}. This perspective is further supported by recent work on the frustrated quantum magnet Cs$_2$RuO$_4$, where similar double-peak features were unambiguously identified in the spin-flop phase \cite{Nabi2025arXiv}.

\section{Conclusion}\label{SEC:CONCLUSION}

In conclusion, we have presented a rich quantum phase diagram containing several intriguing QSLs and symmetry-breaking phases of the spin-$1/2$ Kitaev-$\Gamma$ model in a [111] magnetic field on a honeycomb lattice.
By using ED calculation on a 24-site cluster, we have demonstrated that the $\mathbb{Z}_2$ flux density and TEE are sensitive probes of QPTs.
For the AFM Kitaev model in the external magnetic field, we discover a scalar chiral phase and two vector chiral phases with different ground-state degeneracy at the intermediate region. In particular, the regime of the vector chiral phases shrinks considerably as the $\Gamma$ term increases, implying that careful attention is required to confirm its existence reliably.
For the moderate $\Gamma$ interaction, we identify a proximate QSL characterized by high TEE and three-peak specific heat at a low magnetic field. However, its nature is still elusive and deserves further study.
We also disclose a field-revealed spin-flop phase in a wide parameter range. It is interesting to note that the spin-flop phase exhibits double peaks in its specific heat and has a logarithmic temperature scaling between the two characteristic peaks.
Our work provides a roadmap towards exploring the dynamical and thermal magnetic properties of novel phases in Kitaev antiferromagnets and candidate materials.

\begin{figure}[!ht]
\centering
\includegraphics[width=0.98\columnwidth, clip]{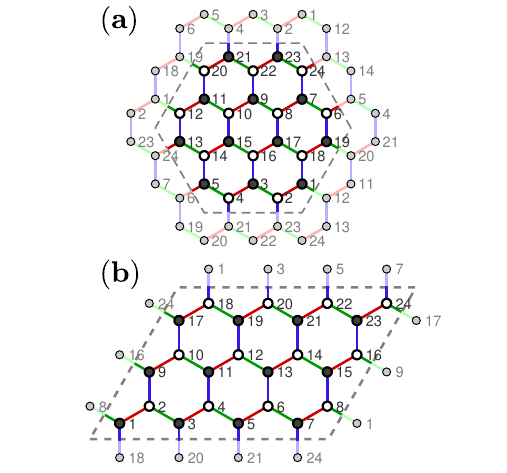}\\
\caption{(a) A 24-site cluster with $C_6$ symmetry under PBCs. (b) A $2 \times 4 \times 3$ torus with PBCs.
The \textbf{X}, \textbf{Y}, and \textbf{Z} bonds are represented by red, green, and blue line segments, respectively. Open circles denote even sites, while filled circles denote odd sites.}
\label{FIG-Lattice}
\end{figure}

\begin{acknowledgements}
We are grateful to Z. Zhou, S. Feng, K. Wang, and H.-Q. Wu for insightful discussions.
Q.L. is supported by the National Natural Science Foundation of China (Grants No. 12304176 and No. 12247183),
the Beijing National Laboratory for Condensed Matter Physics (Grant No. 2025BNLCMPKF022), 
and the Natural Science Foundation of Jiangsu Province (Grant No. BK20220876).
X.W. is supported by the National Program on Key Research Project (Grant No. MOST2022YFA1402701) and the National Natural Science Foundation of China (Grants No. 12574163).
S.L. and H.W. are supported by the Postgraduate Research \& Practice Innovation Program of Jiangsu Province (Grants No. xcxjh20242110 and No. xcxjh20242107). 
The computations are partially supported by High Performance Computing Platform of Nanjing University of Aeronautics and Astronautics. 
Some physical quantities are obtained by employing the open-source exact diagonalization software package $\mathcal{H}\Phi$\cite{Kawamura2017CPC,Ido2024CPC}.
\end{acknowledgements}

\section*{DATA AVAILABILITY}
The data that support the findings of this article are openly available \cite{DataAndCode}.


\appendix

\section{Implementation of PBCs for Lattice and Additional Information on Phase Diagram} \label{SEC:AppA}

To begin with, in Fig. \ref{FIG-Lattice} we present the two lattice geometries employed in our calculations, together with a detailed description of the implementation of PBCs.

\begin{figure}[!ht]
\centering
\includegraphics[width=0.98\columnwidth, clip]{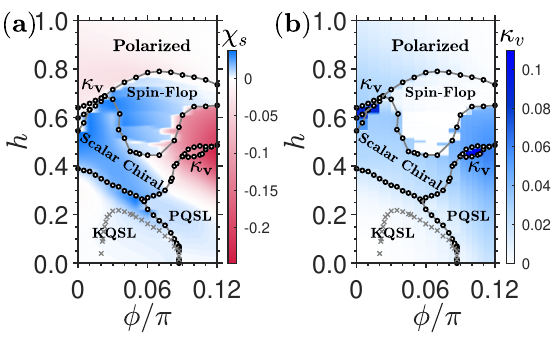}\\
\caption{(a) Phase diagram constructed from the scalar spin chirality defined in Eq.~\eqref{EQ:ScaChiNew}. (b) Phase diagram constructed from the vector spin chirality defined in Eq.~\eqref{EQ:VecChiNew}.}
\label{FIG-ScalarDensity}
\end{figure}

We then show the density plots of the scalar and vector spin chiralities, as illustrated in Fig. \ref{FIG-ScalarDensity}. It should be noted that, due to limitations in computational resources, the spatial resolution of these density maps is finite.
As shown in Fig. \ref{FIG-ScalarDensity}(a), the phase boundaries between the KQSL and the scalar chiral phase, between the scalar chiral and vector chiral phases, as well as those separating the spin-flop and polarized phases, can be relatively easily identified. However, the boundary between the scalar chiral phase and the spin-flop phase is difficult to distinguish solely using the scalar chirality order parameter. Moreover, in both the scalar chiral and PQSL phases, the scalar chirality order parameter exhibits both positive and negative values, indicating a pronounced spatial inhomogeneity.
In Fig. \ref{FIG-ScalarDensity}(b), the vector chiral phase clearly displays large vector chirality order parameters, allowing for a clear identification of the phase boundaries between the KQSL and the scalar chiral and PQSL phases. The spin-flop and vector chiral phases can also be readily distinguished from other phases. Nevertheless, the vector chirality order parameter remains insufficient for distinguishing between the scalar chiral and PQSL phases. This suggests that the PQSL phase may also host finite scalar and/or vector chiralities, rendering it difficult to accurately map out the full phase diagram using scalar or vector chirality order parameters alone.

\begin{figure}[!ht]
\centering
\includegraphics[width=0.95\columnwidth, clip]{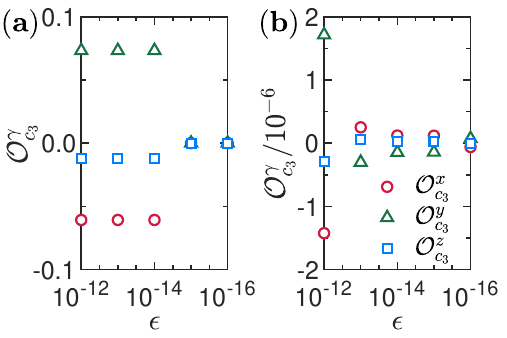}\\
\caption{Illustration of the convergence behavior of the order parameters $\mathcal{O}_{C_{3}}^{x}$ (red circles), $\mathcal{O}_{C_{3}}^{y}$ (green triangles), and $\mathcal{O}_{C_{3}}^{z}$ (blue squares) with respect to the energy convergence tolerance $\epsilon$ of (a) the first and (b) the third energy level at the representative point $(\phi/\pi = 0.01, h = 0.01)$.}
\label{FIG-BondEn}
\end{figure}

\section{Spurious Lattice Nematicity in Low-Field KQSL}\label{SEC:AppB}

The lattice nematicity results from the breaking of lattice-rotational symmetry and can be detected through differences in the bond energy $E_{\gamma} = \big\langle \mathcal{H} \big\rangle_{\langle i,j\rangle_{\gamma}}$, where $\gamma \in {x,y,z}$ denotes the bond type. When $C_3$ symmetry is broken, the bond energies $E_{\gamma}$ become distinct across different bond types. To quantify this, we define a $C_3$-symmetry-breaking order parameter for each bond type $\gamma$ as $\mathcal{O}_{C_3}^{\gamma}$, which measures the deviation of $E_{\gamma}$ from the average bond energy of the system \cite{Gohlke2020PRR}. Explicitly, for each $\gamma$, the order parameter is given by
\begin{align}\label{EQ:LattNematic}
\mathcal{O}_{C_{3}}^{\gamma} = E_{\gamma} - \frac{1}{3} \big( E_{x} + E_{y} + E_{z} \big).
\end{align}

In the low-field KQSL, the ground state exhibits quasi-threefold degeneracy. Due to the significantly slower convergence of eigenstates compared to eigenvalues, the values of the nematic order parameters $\mathcal{O}_{C_{3}}^{\gamma}$ vary considerably depending on the convergence criterion set by the numerical order of the energy levels.
Taking the representative point $(\phi/\pi, h) = (0.01, 0.01)$ as an example, we examine the convergence behavior of the three nematic order parameters $\mathcal{O}_{C_{3}}^{\gamma}$ with respect to the convergence tolerance of the first [Fig.~\ref{FIG-BondEn}(a)] and the third energy level [Fig.~\ref{FIG-BondEn}(b)]. As the numerical precision approaches the machine epsilon (approximately $2.22 \times 10^{-16}$ in C++ using double-precision arithmetic), the values of $\mathcal{O}_{C_{3}}^{\gamma}$ remain finite in Fig.~\ref{FIG-BondEn}(a), whereas they gradually decrease and become vanishingly small in Fig.~\ref{FIG-BondEn}(b).
These results indicate that the observed lattice nematicity in the low-field KQSL is not physical, but rather a spurious effect arising from ground-state degeneracy and finite numerical precision.

\section{Evolution and Instability of Intermediate Phases with Field Tilt} \label{SEC:AppC}

As revealed by our high-precision and high-resolution ED calculation of the AFM Kitaev model in a [111] magnetic field on the $C_{6}$-symmetric 24-site cluster [see Fig. ~\ref{FIG-Lattice}(a)], the intermediate region, which is sandwiched between the KQSL and the polarized phase, contains three different phases. Among these, apart from a wide scalar chiral phase, we also identify two vector chiral phases with distinct ground-state degeneracy.
In contrast, in a similar ED study by Hickey and Trebst~\cite{Hickey2019NC}, they discovered the emergence of a $U(1)$ QSL in the nearly same region of the scalar chiral phase. Nonetheless, they did not report the existence of vector chiral phases. In addition, they found that the intermediate phase survives against the tilted magnetic fields at least for small tilt angles.
In fact, although the phase diagram in Ref.~\cite{Hickey2019NC} encompasses the $[111]$ direction, the main results presented in that work—such as flux density $\overline{W}_{\!p}$, ground-state fidelity, and the second derivative of the energy—were computed for representative directions tilted away from $\textbf{c} [111]$ axis towards $\textbf{b} [\bar110]$ axis with different deviation angles such as $\vartheta = 7.5^\circ$.

\begin{figure}[!ht]
\centering
\includegraphics[width=0.95\columnwidth, clip]{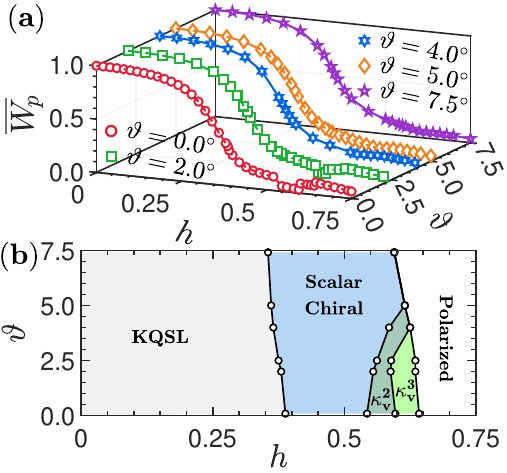}\\
\caption{(a) Flux density $\overline{W}_{\!p}$ versus magnetic field strength $h$ for various deviation angles $\vartheta$. The angle $\vartheta$ is defined as the tilt of the magnetic field away from the $c$-axis within the $bc$ plane. (b) Phase diagram in the parameter space of $h$ and $\vartheta$. The regions labeled $\mathbf{\kappa_v^2}$ and $\mathbf{\kappa_v^3}$ correspond to vector chiral phases with ground-state degeneracies of two and three, respectively.}
\label{FIG-AnglePhase}
\end{figure}

To systematically investigate this field-directional dependence, we performed a high-resolution scan of the magnetic field orientation within the $\textbf{b} [\bar110]$–$\textbf{c} [111]$ plane, varying the deviation angle from the $c$-axis continuously from $0^\circ$ to $7.5^\circ$. Multiple diagnostic quantities including the flux density $\overline{W}_{\!p}$ were computed concurrently. The results reveal that the intermediate-field region exhibits a strong dependence on the deviation angle: it splits into three distinct phases for angles below approximately $4.0^\circ$, merges into two phases within the $4.0^\circ$–$5.0^\circ$ range, and finally collapses into a single phase once the deviation exceeds about $5.0^\circ$. The evolution of $\overline{W}_{\!p}$ with magnetic field strength across different deviation angles is illustrated in Fig.~\ref{FIG-AnglePhase}~(a), while the corresponding phase diagram in the $(h, \vartheta)$ plane is presented in Fig.~\ref{FIG-AnglePhase}~(b).
Hence, our work reveals that in the immediate vicinity of the [111] direction, the ground state exhibits extreme sensitivity to tiny deviations and hosts a more complex multiphase structure. This outcome clarifies seemingly inconsistent findings in the literature and underscores the vital importance of precise control and theoretical examination of magnetic field orientation when exploring the precise phase diagram of Kitaev materials.

\section{Unveiling an Overlooked Narrow Phase on the $2\times4\times3$ Torus}\label{SEC:AppD}

In addition to the $C_{6}$-symmetric 24-site cluster, there are also several ED studies of the AFM Kitaev model in a [111] magnetic field on the $2\times4\times3$ torus,
as depicted in Fig.~\ref{FIG-Lattice}(b).
These studies coincidentally identify a sole intermediate phase.
However, in contrast to the low-field QPT at $h_t^l \approx 0.365$, the high-field QPT near $h_t^h \approx 0.620$, as identified via magnetic susceptibility \cite{Zhu2018PRB} or topological entanglement entropy \cite{Ronquillo2019PRB}, appears somewhat broadened, suggesting the potential for a narrow intervening phase. 
For the torus geometry considered here, Wilson loop operators provide a convenient and efficient means of detecting QPTs. Thus, the purpose of this Appendix is to investigate the possible existence of such an additional phase.

\begin{figure}[!ht]
\centering
\includegraphics[width=0.95\columnwidth, clip]{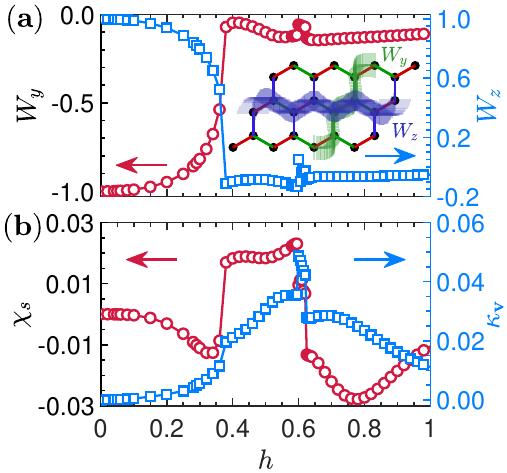}\\
\caption{(a) $W_y$ (left axis, red circles) and $W_z$ (right axis, blue squares) as a function of the magnetic field $h$. The inset depicts the construction of $W_y$ and $W_z$. (b) The scalar chiral order parameter $\chi_s$ (left axis) and the vector chiral order parameter $\kappa_v$ (right axis) as a function of $h$.}
\label{FIG-WyWz_Chiralities}
\end{figure}

As shown in the inset of Fig.~\ref{FIG-WyWz_Chiralities}(a), the Wilson loop operators $W_y$ and $W_z$ are defined as
\begin{align}\label{EQ:WilsonLoop}
W_y = - \Big\langle \prod_{i=1}^{2L_y} \sigma_i^y \Big\rangle, \ 
W_z = - \Big\langle \prod_{i=1}^{2L_z} \sigma_i^z \Big\rangle.
\end{align}
where $\sigma_y$ and $\sigma_z$ are Pauli matrices satisfying $\mathbf{\hat S} = \frac{\hbar}{2}\mathbf{\hat\sigma} = \frac{\hbar}{2}(\sigma_x, \sigma_y, \sigma_z)$. 
In the zero-field Kitaev limit, it can be readily verified that these Wilson loop operators commute with the Hamiltonian, and their eigenvalues take values of $\pm 1$ \cite{Zhu2018PRB}.
As indicated by the red circles in Fig.~\ref{FIG-WyWz_Chiralities}(a), $W_y$ starts from an initial value of $-1$, exhibits the first abrupt change at $h_t^l \approx 0.365$, followed by a second abrupt change at $h_t^m \approx 0.595$, and finally enters the polarized phase at $h_t^h \approx 0.620$. Similar abrupt changes are observed in the $W_z$ curve [blue squares in Fig.~\ref{FIG-WyWz_Chiralities}(a)], confirming the existence of two intermediate phases in the range of $h\in(0.365,0.595)$ and $(0.595,0.620)$.
We note that the Wilson loop calculation by Zhu \textit{et al.} \cite{Zhu2018PRB}, which is limited to fields up to $h \simeq 0.5$, places the low-field QPT at $h_t^l \approx 0.2\sqrt{3} \approx 0.346$, a value consistent with our result.
To clarify the nature of these two phases, we computed the scalar chiral order parameter ($\chi_s$) and the vector chiral order parameter ($\kappa_v$).
It is observed from Fig.~\ref{FIG-WyWz_Chiralities}(b) that $\chi_s$ and $\kappa_v$ are pronounced in the first and second intermediate phases, respectively.
Our results thus highlight the existence of a narrow vector chiral phase which was overlooked previously.
This omission may be attributed to the notably narrow extent of the phase and the restricted resolution of previously employed diagnostics, highlighting the challenge in detecting delicate yet potentially crucial states in quantum phase diagrams.


%






\begin{thebibliography}{99}%

\bibitem{Starykh2015RPP}
   O. A. Starykh,
   Unusual ordered phases of highly frustrated magnets: a review,
   \href{ https://dx.doi.org/10.1088/0034-4885/78/5/052502}{Rep. Prog. Phys. \textbf{78}, 052502 (2015)}.

\bibitem{Cui2023FoP}
   Q. Cui, L. Wang, Y. Zhu, J. Liang, and H. Yang,
   Magnetic anisotropy, exchange coupling and Dzyaloshinskii-Moriya interaction of two-dimensional magnets,
   \href{ https://dx.doi.org/10.1007/s11467-022-1217-7}{Front. Phys. \textbf{18}, 13602 (2023)}.
   
\bibitem{Rousochatzakis2024RPP}
   I. Rousochatzakis, N. B. Perkins, Q. Luo, and H.-Y. Kee,
   Beyond Kitaev physics in strong spin-orbit coupled magnets,
   \href{ https://dx.doi.org/10.1088/1361-6633/ad208d}{Rep. Prog. Phys. \textbf{87}, 026502 (2024)}.
   
\bibitem{Luo2025CPL}
  Q. Luo, J. Zhao, and X. Wang,
  Interplay of Kitaev Interaction and Off-Diagonal Exchanges: Exotic Phases and Quantum Phase Diagrams,
  \href{https://doi.org/10.1088/0256-307X/42/2/027501}{Chin. Phys. Lett. \textbf{42}, 027501 (2025).}

\bibitem{Hu2020NC}
    Z. Hu, Z. Ma, Y.-D. Liao, H. Li, C. Ma, Y. Cui, Y. Shangguan, Z. Huang, Y. Qi, W. Li, Z. Y. Meng, J. Wen, and W. Yu,
    Evidence of the Berezinskii-Kosterlitz-Thouless Phase in a Frustrated Magnet,
    \href{https://doi.org/10.1038/s41467-020-19380-x}{Nat. Commun. \textbf{11}, 5631 (2020)}.
   
\bibitem{Li2020NC}
   H. Li, Y. D. Liao, B.-B. Chen, X.-T. Zeng, X.-L. Sheng, Y. Qi, Z. Y. Meng , and W. Li,
   Kosterlitz-Thouless melting of magnetic order in the triangular quantum Ising material TmMgGaO$_4$,
   \href{https://doi.org/10.1038/s41467-020-14907-8}{Nat. Commun. \textbf{11}, 1111 (2020)}.

\bibitem{Muhlbaue2009science}
  S. M\"{u}hlbauer, B. Binz, F. Jonietz, C. Pfleiderer, A. Rosch, A. Neubauer, R. Georgii, and P. B\"{o}ni,
  Skyrmion Lattice in a Chiral Magnet,
  \href{https://doi.org/10.1126/science.1166767}{Science \textbf{323}, 915-919(2009)}.

\bibitem{Yi2009PRB}
   S. D. Yi, S. Onoda, N. Nagaosa, and J. H. Han,
   Skyrmions and anomalous Hall effect in a Dzyaloshinskii-Moriya spiral magnet,
   \href{ https://doi.org/10.1103/PhysRevB.80.054416}{Phys. Rev. B \textbf{80}, 054416 (2009)}.

\bibitem{Baek2017PRL}
   S.-H. Baek, S.-H. Do, K.-Y. Choi, Y. S. Kwon, A. U. B. Wolter, S. Nishimoto, J. van den Brink, and B. B\"{u}chner, 
   Evidence for a Field-Induced Quantum Spin Liquid in $\alpha$-Rucl$_{3}$,
   \href{ https://doi.org/10.1103/PhysRevLett.119.037201}{Phys. Rev. Lett. \textbf{119}, 037201 (2017)}.

\bibitem{Leahy2017PRL}
   I. A. Leahy, C. A. Pocs, P. E. Siegfried, D. Graf, S.-H. Do, K.-Y. Choi, B. Normand, and M. Lee, 
   Anomalous Thermal Conductivity and Magnetic Torque Response in the Honeycomb Magnet $\alpha$-Rucl$_{3}$,
   \href{ https://doi.org/10.1103/PhysRevLett.118.187203}{Phys. Rev. Lett. \textbf{118}, 187203 (2017)}.

\bibitem{Zheng2017PRL}
   J. Zheng, K. Ran, T. Li, J. Wang, P.-S. Wang, B. Liu, Z.-X. Liu, B. Normand, J. Wen, and W. Yu,
   Anomalous Thermal Conductivity and Magnetic Torque Response in the Honeycomb Magnet $\alpha$-RuCl$_{3}$,
   \href{ https://doi.org/10.1103/PhysRevLett.119.227208}{Phys. Rev. Lett. \textbf{119}, 227208 (2017)}.

\bibitem{Banerjee2018npj}
  A. Banerjee, P. Lampen-Kelley, J. Knolle, C. Balz, A. A. Aczel, B. Winn, Y. Liu, D. Pajerowski, J. Yan, C. A. Bridges, A. T. Savici, B. C. Chakoumakos, M. D. Lumsden, D. A. Tennant, R. Moessner, D. G. Mandrus, and S. E. Nagler,
  Excitations in the field-induced quantum spin liquid state of $\alpha$-RuCl$_{3}$,
  \href{https://doi.org/10.1038/s41535-018-0079-2}{npj Quantum Mater. \textbf{3}, 8 (2018)}.

\bibitem{Zhou2023NC}
  X.-G. Zhou, H. Li, Y. H. Matsuda, A. Matsuo, W. Li, N. Kurita, G. Su, K. Kindo, and H. Tanaka, 
  Possible Intermediate Quantum Spin Liquid Phase in $\alpha$-RuCl$_3$ under High Magnetic Fields up to 100 T,
  \href{https://doi.org/10.1038/s41467-023-41232-7}{Nat. Commun. \textbf{14}, 5613 (2023)}.
  
\bibitem{Kitaev2006Ann.Phys}
  A. Kitaev, 
  Anyons in an exactly solved model and beyond,
  \href{ https://doi.org/10.1016/j.aop.2005.10.005}{Ann. Phys. (NY) \textbf{321}, 2 (2006)}.

\bibitem{Jackeli2009PRL}
   G. Jackeli, and G. Khaliullin, 
   Mott Insulators in the Strong Spin-Orbit Coupling Limit: From Heisenberg to a Quantum Compass and Kitaev Models,
   \href{ https://doi.org/10.1103/PhysRevLett.102.017205}{Phys. Rev. Lett. \textbf{102}, 017205 (2009)}.

\bibitem{Ye2012PRB}
   F. Ye, S. X. Chi, H. B. Cao, B. C. Chakoumakos, J. A. Fernandez-Baca, R. Custelcean, T. F. Qi, O. B. Korneta, and G. Cao, 
   Direct evidence of a zigzag spin-chain structure in the honeycomb lattice: A neutron and x-ray diffraction investigation of single-crystal Na$_{2}$IrO$_{3}$,
   \href{ https://doi.org/10.1103/PhysRevB.85.180403}{Phys. Rev. B. \textbf{85}, 180403 (2012)}.

\bibitem{Choi2012PRL}
   S. K. Choi, R. Coldea, A. N. Kolmogorov, T. Lancaster, I. I. Mazin, S. J. Blundell, P. G. Radaelli, Y. Singh, P. Gegenwart, K. R. Choi, S.-W. Cheong, P. J. Baker, C. Stock, and J. Taylor, 
   Spin Waves and Revised Crystal Structure of Honeycomb Iridate Na$_{2}$IrO$_{3}$,
   \href{ https://doi.org/10.1103/PhysRevLett.108.127204}{Phys. Rev. Lett. \textbf{108}, 127204 (2012)}.

\bibitem{Rau2014PRL}
   J. G. Rau, E. K.-H. Lee, and H.-Y. Kee, 
   Generic Spin Model for the Honeycomb Iridates beyond the Kitaev Limit,
   \href{ https://doi.org/10.1103/PhysRevLett.112.077204}{Phys. Rev. Lett. \textbf{112}, 077204 (2014)}.


   
\bibitem{Sears2015PRB}
   J. A. Sears, M. Songvilay, K. W. Plumb, J. P. Clancy, Y. Qiu, Y. Zhao, D. Parshall, and Y.-J. Kim,
   Magnetic order in $\alpha$-Rucl$_{3}$: A honeycomb-lattice quantum magnet with strong spin-orbit coupling,
   \href{ https://doi.org/10.1103/PhysRevB.91.144420}{Phys. Rev. B \textbf{91}, 144420 (2015)}.

\bibitem{Kim2015PRB}
   H.-S. Kim, V. S. V., A. Catuneanu, and H.-Y. Kee,
   Kitaev magnetism in honeycomb RuCl$_{3}$ with intermediate spin-orbit coupling,
   \href{    https://doi.org/10.1103/PhysRevB.91.241110}{Phys. Rev. B \textbf{91}, 241110(R) (2015)}.
   
\bibitem{Johnson2015PRB}
   R. D. Johnson, S. C. Williams, A. A. Haghighirad, J. Singleton, V. Zapf, P. Manuel, I. I. Mazin, Y. Li, H. O. Jeschke, R. Valent{\'l}, and R. Coldea,
   Magnetic order in $\alpha$-RuCl$_{3}$: A honeycomb-lattice quantum magnet with strong spin-orbit coupling,
   \href{ https://doi.org/10.1103/PhysRevB.92.235119}{Phys. Rev. B \textbf{92}, 235119 (2015)}.

\bibitem{Liu2018PRB}
    H. Liu, and G. Khaliullin,
    Pseudospin exchange interactions in $d^7$ cobalt compounds: Possible realization of the Kitaev model,
   \href{https://doi.org/10.1103/PhysRevB.97.014407}{Phys. Rev. B \textbf{97}, 014407 (2018)}.

\bibitem{Yao2020PRB}
  W. Yao and Y. Li,
  Ferrimagnetism and anisotropic phase tunability by magnetic fields in Na$_2$Co$_2$TeO$_6$,
  \href{https://doi.org/10.1103/PhysRevB.101.085120}{Phys. Rev. B \textbf{101}, 085120 (2020)}.

\bibitem{Lin2021NC}
   G. Lin, J. Jeong, C. Kim, Y. Wang, Q. Huang, T. Masuda, S. Asai, S. Itoh, G. G\"{u}nther, M. Russina, Z. Lu, J. Sheng, L. Wang, J. Wang, G. Wang, Q. Ren, C. Xi, W. Tong, L. Ling, Z. Liu, L. Wu, J. Mei, Z. Qu, H. Zhou, X. Wang, J.-G. Park, Y. Wan, and J. Ma,
   Field-induced quantum spin disordered state in spin-1/2 honeycomb magnet Na$_2$Co$_2$TeO$_6$,
   \href{https://doi.org/10.1038/s41467-021-25567-7}{Nat. Commun. \textbf{12}, 5559 (2021)}. 

\bibitem{Li2022PRX}
   X. Li, Y. Gu, Y. Chen, V. O. Garlea, K. Iida, K. Kamazawa, Y. Li, G. Deng, Q. Xiao, X. Zheng, Z. Ye, Y. Peng, I. A. Zaliznyak, J. M. Tranquada, and Y. Li,
   Giant Magnetic In-Plane Anisotropy and Competing Instabilities in Na$_3$Co$_2$SbO$_6$,
   \href{https://doi.org/10.1103/PhysRevX.12.041024}
   {Phys. Rev. X \textbf{12}, 041024 (2022)}.

\bibitem{Jiao2024CM}
    J. Jiao, X. Li, G. Lin, M. Shu, W. Xu, O. Zaharko, T. Shiroka, T. Hong, A. I. Kolesnikov, G. Deng, S. Dunsiger, M. C. Aronson, H. Zhou, X. Wang, T. Shang, and J. Ma,
    Static magnetic order with strong quantum fluctuations in spin-1/2 honeycomb magnet Na$_2$Co$_2$TeO$_6$,
   \href{https://doi.org/10.1038/s43246-024-00594-1}{Commun Mater \textbf{5}, 159 (2024)}.

\bibitem{Sugita2020PRB}
   Y. Sugita, Y. Kato, and Y. Motome,
   Antiferromagnetic Kitaev interactions in polar spin-orbit Mott insulators,
   \href{ https://doi.org/10.1103/PhysRevB.101.100410}{Phys. Rev. B \textbf{101}, 100410(R) (2020)}.

\bibitem{Xia2025arXiv}
   L.-H. Xia, Y.-P. Gao, Z.-Y. Dong, and J.-X. Li,
   Ligand-SOC enhanced $4f^{5}$ Kitaev antiferromagnet: Application to $\mathrm{SmI_{3}}$,
   \href{ https://doi.org/10.48550/arXiv.2505.18616}{arXiv:2505.18616}.

\bibitem{Jiang2011PRB}
   H.-C. Jiang, Z.-C. Gu, X.-L. Qi, and S. Trebst,
   Possible proximity of the Mott insulating iridate Na$_2$IrO$_3$ to a topological phase: Phase diagram of the Heisenberg-Kitaev model in a magnetic field,
   \href{  https://doi.org/10.1103/PhysRevB.83.245104}{Phys. Rev. B \textbf{83}, 245104 (2011)}.

\bibitem{Zhu2018PRB}
   Z. Zhu, I. Kimchi, D. N. Sheng, and L. Fu,
   Robust non-Abelian spin liquid and a possible intermediate phase in the antiferromagnetic Kitaev model with magnetic field,
   \href{  https://doi.org/10.1103/PhysRevB.97.241110}{Phys. Rev. B \textbf{97}, 241110(R) (2018)}.

\bibitem{Hickey2019NC}
   C. Hickey, and S. Trebst,
   Emergence of a field-driven U(1) spin liquid in the Kitaev honeycomb model,
   \href{https://doi.org/10.1038/s41467-019-08459-9}{Nat. Commun. \textbf{10}, 530 (2019)}.

\bibitem{Ronquillo2019PRB}
  D. C. Ronquillo, A. Vengal, and N. Trivedi,
  Signatures of magnetic-field-driven quantum phase transitions in the entanglement entropy and spin dynamics of the Kitaev honeycomb model,
  \href{ https://doi.org/10.1103/PhysRevB.99.140413}{Phys. Rev. B \textbf{99}, 140413(R) (2019)}.


\bibitem{Gohlke2018PRB}
   M. Gohlke, R. Moessner, and F. Pollmann,
   Dynamical and topological properties of the Kitaev model in a [111] magnetic field,
   \href{ https://doi.org/10.1103/PhysRevB.98.014418}{Phys. Rev. B \textbf{98}, 014418 (2018)}.


\bibitem{Patel2019pnas}
  N. D. Patel, and N. Trivedi,
  Magnetic ﬁeld-induced intermediate quantum spinliquid with a spinon Fermi surface,
  \href{https://doi.org/10.1073/pnas.1821406116}{Proc. Natl. Acad. Sci. USA \textbf{116}, 12199 (2019)}.


\bibitem{Zhang2022NC}
  S.-S. Zhang, G. B. Hal\'{a}sz, and C. D. Batista,
  Theory of the Kitaev model in a [111] magnetic field,
  \href{https://doi.org/10.1038/s41467-022-28014-3}{Nat. Commun. \textbf{13}, 399 (2022)}.

\bibitem{Wang2025PRB}
   K. Wang, S. Feng, P. Zhu, R. Chi, H.-J. Liao, N. Trivedi, and T. Xiang, 
   Fractionalization signatures in the dynamics of quantum spin liquids,
   \href{ https://doi.org/10.1103/PhysRevB.111.L100402}{Phys. Rev. B \textbf{111}, L100402 (2025)}.

\bibitem{Zhu2025NC}
  P. Zhu, S. Feng, K. Wang, T. Xiang, and N. Trivedi, 
  Emergent quantum Majorana metal from a chiral spin liquid,
   \href{https://doi.org/10.1038/s41467-025-56789-8}{Nat. Commun. \textbf{16}, 2420 (2025)}.

\bibitem{Jiang2020PRL}
   M.-H. Jiang, S. Liang, W. Chen, Y. Qi, J.-X. Li, and Q.-H. Wang, 
   Tuning Topological Orders by a Conical Magnetic Field in the Kitaev Model,
   \href{https://doi.org/10.1103/PhysRevLett.125.177203}{Phys. Rev. B \textbf{125}, 177203 (2020)}.

\bibitem{Chen2025PRB}
   C. Chen and I. S. Villadiego, 
   Anyon polarons as a window into competing phases of the Kitaev honeycomb model under a Zeeman field,
   \href{ https://doi.org/10.1103/xxym-pc1t}{Phys. Rev. B \textbf{111}, 245140 (2025)}.

\bibitem{Holdhusen2024PRB}
   W. Holdhusen, D. Huerga, and G. Ortiz, 
   Emergent magnetic order in the antiferromagnetic Kitaev model in a [111] field,
   \href{ https://doi.org/10.1103/PhysRevB.109.174411}{Phys. Rev. B \textbf{109}, 174411 (2024)}.
   
\bibitem{Matsuda2025arXiv}
  Y. Matsuda, T. Shibauchi, H.-Y. Kee,
  Kitaev Quantum Spin Liquids,
  \href{https://arxiv.org/abs/2501.05608}{arXiv:2501.05608}.

\bibitem{Wang2017PRB}
  W. Wang, Z.-Y. Dong, S.-L. Yu, and J.-X. Li,
  Theoretical investigation of magnetic dynamics in $\alpha$-RuCl$_3$,
  \href{https://doi.org/10.1103/PhysRevB.96.115103}{Phys. Rev. B \textbf{96}, 115103 (2017).}
  
\bibitem{Gohlke2020PRR}
   M. Gohlke, L. E. Chern, H.-Y. Kee, and Y. B. Kim,
   Emergence of nematic paramagnet via quantum order-by-disorder and pseudo-Goldstone modes in Kitaev magnets,
   \href{ https://doi.org/10.1103/PhysRevResearch.2.043023}{Phys. Rev. Research \textbf{2}, 043023 (2020)}.

\bibitem{Wang2019PRL}
  J. Wang, B. Normand, and Z.-X. Liu,
  One Proximate Kitaev Spin Liquid in the $K$-$J$-$\Gamma$ Model on the Honeycomb Lattice,
  \href{https://doi.org/10.1103/PhysRevLett.123.197201}{Phys. Rev. Lett. \textbf{123}, 197201 (2019).}

\bibitem{Lee2020NC}
   H.-Y. Lee, R. Kaneko, L. E. Chern, T. Okubo, Y. Yamaji, N. Kawashima, and Y. B. Kim,
   Magnetic field induced quantum phases in a tensor network study of Kitaev magnets,
   \href{https://doi.org/10.1038/s41467-020-15320-x}{Nat. Commun. \textbf{11}, 1639 (2020)}.

\bibitem{Luo2024PRB}
   Q. Luo, J. Zhao, J. Li, and X. Wang,
   Chiral spin state and nematic ferromagnet in the spin-1 Kitaev-model,
   \href{ https://doi.org/10.1103/PhysRevB.110.035121}{Phys. Rev. B \textbf{110}, 035121 (2024)}.

\bibitem{Zou2025PRB}
   Z. Zou and Q. Luo,
   High Chern numbers and topological flat bands in high-field polarized Kitaev magnets on the star lattice,
   \href{ https://doi.org/10.1103/wf94-z249}{Phys. Rev. B \textbf{112}, 054409 (2025)}.

\bibitem{Rousochatzakis2023KITP}
  I. Rousochatzakis,
  Interplay between classical and quantum spin liquids in the spin-$1/2$ Kitaev-Gamma model on the honeycomb lattice,
  \href{https://doi.org/10.26081/K6XX0J}{KITP Program: A New Spin on Quantum Magnets (Jul. 17 - Sep. 8, 2023)}.
  
\bibitem{Yang2020PRL}
  W. Yang, A. Nocera, T. Tummuru, H.-Y. Kee, and I. Affleck,
  Phase Diagram of the Spin-$1/2$ Kitaev-Gamma Chain and Emergent SU(2) Symmetry,
  \href{https://doi.org/10.1103/PhysRevLett.124.147205}{Phys. Rev. Lett. \textbf{124}, 147205 (2020)}.

\bibitem{Luo2021PRB}
  Q. Luo, J. Zhao, X. Wang, and H.-Y. Kee,
  Unveiling the phase diagram of a bond-alternating spin-$\frac12$ $K$-$\Gamma$ chain,
  \href{https://doi.org/10.1103/PhysRevB.103.144423}{Phys. Rev. B \textbf{103}, 144423 (2021)}.

\bibitem{Luo2021PRR}
   Q. Luo, S. Hu, and H.-Y. Kee, 
   Unusual excitations and double-peak specific heat in a bond-alternating spin-1 chain,
   \href{  https://doi.org/10.1103/PhysRevResearch.3.033048}{Phys. Rev. Research \textbf{3}, 033048 (2021)}.

\bibitem{Sorensen2021PRX}
  E. S. S{\o}rensen, A. Catuneanu, J. S. Gordon, and H.-Y. Kee, 
  Heart of Entanglement: Chiral, Nematic, and Incommensurate Phases in the Kitaev-$\Gamma$ Ladder in a Field,
  \href{ https://doi.org/10.1103/PhysRevX.11.011013}{Phys. Rev. X \textbf{11}, 011013 (2021)}.
  
\bibitem{Sorensen2024npj}
   E. S. S{\o}rensen, and H.-Y. Kee, 
   Twice hidden string order and competing phases in the spin-1/2 Kitaev–Gamma ladder,
   \href{https://doi.org/10.1038/s41535-024-00621-x}{npj Quantum Mater. \textbf{9}, 10 (2024)}.
   
\bibitem{Luo2021NPJ}
  Q. Luo, J. Zhao, H.-Y. Kee, and X. Wang,
  Gapless quantum spin liquid in a honeycomb $\Gamma$ magnet,
  \href{https://doi.org/10.1038/s41535-021-00356-z}{npj Quantum Mater. \textbf{6}, 57 (2021).}

\bibitem{Feng2023PRB}
   S. Feng, A. Agarwala, S. Bhattacharjee, and N. Trivedi,
   Anyon dynamics in field-driven phases of the anisotropic Kitaev model,
  \href{ https://doi.org/10.1103/PhysRevB.108.035149}{Phys. Rev. B \textbf{108}, 035149 (2023)}.

\bibitem{Kitaev2006PRL}
   A. Kitaev, and J. Preskill,
   Topological Entanglement Entropy,
   \href{ https://doi.org/10.1103/PhysRevLett.96.110404}{Phys. Rev. Lett. \textbf{96}, 110404 (2006)}.
   
\bibitem{Knolle2014PRL}
   J. Knolle, D. L. Kovrizhin, J. T. Chalker, and R. Moessner,
   Dynamics of a Two-Dimensional Quantum Spin Liquid: Signatures of Emergent Majorana Fermions and Fluxes,
   \href{ https://doi.org/10.1103/PhysRevLett.112.207203}{Phys. Rev. Lett. \textbf{112}, 207203 (2014)}.
   
\bibitem{Knolle2015PRB}
   J. Knolle, D. L. Kovrizhin, J. T. Chalker, and R. Moessner,
   Dynamics of fractionalization in quantum spin liquids,
  \href{ https://doi.org/10.1103/PhysRevB.92.115127}{Phys. Rev. B \textbf{92}, 115127 (2015)}.

\bibitem{Nasu2015PRB}
   J. Nasu, M. Udagawa, and Y. Motome,
   Thermal fractionalization of quantum spins in a Kitaev model: Temperature-linear specific heat and coherent transport of Majorana fermions,
  \href{ https://doi.org/10.1103/PhysRevB.92.115122}{Phys. Rev. B \textbf{92}, 115122 (2015)}.

\bibitem{Kurita2015PRB}
   M. Kurita, Y. Yamaji, S. Morita, and M. Imada, 
   Variational Monte Carlo method in the presence of spin-orbit interaction and its application to Kitaev and Kitaev-Heisenberg models,
   \href{ https://doi.org/10.1103/PhysRevB.92.035122}{Phys. Rev. B \textbf{92}, 035122 (2015)}.
  
\bibitem{Sugiura2013PRL}
   S. Sugiura, and A. Shimizu,
   Canonical Thermal Pure Quantum State,
  \href{https://doi.org/10.1103/PhysRevLett.111.010401}{Phys. Rev. Lett. \textbf{111}, 010401 (2013)}.
  
\bibitem{Kawamura2017CPC}
   M. Kawamura, K. Yoshimi, T. Misawa, Y. Yamaji, S. Todo, and N. Kawashima,
   Quantum lattice model solver $\mathcal{H}\phi$,
  \href{https://doi.org/10.1016/j.cpc.2017.04.006}{Comput. Phys. Commun. \textbf{217}, 180 (2017)}.

\bibitem{Ido2024CPC}
   K. Ido, M. Kawamura, Y. Motoyama, K. Yoshimi, Y. Yamaji, S. Todo, N. Kawashima, and T. Misawa,
   Update of $\mathcal{H}\phi$ : Newly added functions and methods in versions 2 and 3r ,
  \href{https://doi.org/10.1016/j.cpc.2024.109093}{Comput. Phys. Commun. \textbf{298}, 109093 (2024)}.
   
\bibitem{Luo2022PRR}
   Q. Luo, P. P. Stavropoulos, J. S. Gordon, and H.-Y. Kee, 
   Spontaneous chiral-spin ordering in spin-orbit coupled honeycomb magnets,
   \href{ https://doi.org/10.1103/PhysRevResearch.4.013062}{Phys. Rev. Research \textbf{4}, 013062 (2022)}.

\bibitem{Laurell2020npj}
   P. Laurell, and S. Okamoto, 
   Dynamical and thermal magnetic properties of the Kitaev spin liquid candidate $\alpha$-RuCl$_{3}$,
   \href{https://doi.org/10.1038/s41535-019-0203-y}{npj Quantum Mater. \textbf{5}, 2 (2020)}.
   
\bibitem{Gagliano1987PRL}
   E. R. Gagliano, and C. A. Balseiro, 
   Dynamical Properties of Quantum Many-Body Systems at Zero Temperature,
   \href{ https://doi.org/10.1103/PhysRevLett.59.2999}{Phys. Rev. Lett. \textbf{59}, 2999 (1987)}.

\bibitem{Yu2024PRB}
   Z. Yu, H. Ding, K. Zhai, C. Mu, A. Nie, J. Cong, J. Huang, H. Zhou, Q. Wang, F. Wen, J. Xiang, B. Wang, T. Xue, Z. Zeng, and Z. Liu,
   Crystal structure, spin flop transition, and magnetoelectric effect in the honeycomb-lattice frustrated Fe-doped Ni$_{2}$Mo$_{3}$O$_{8}$ antiferromagnets,
  \href{https://doi.org/10.1103/PhysRevB.109.024442}{Phys. Rev. B \textbf{109}, 024442 (2024)}.

\bibitem{Huang2025PRB}
   X. Huang, P. Ma, M. Huo, C. Huang, H. Sun, X. Chen, S. Deng, L. He, Tao Xie, Z. He, and M. Wang,
   Ground state and magnetic transitions of the orthorhombic antiferromagnet CaCo$_{2}$TeO$_{6}$,
  \href{ https://doi.org/10.1103/PhysRevB.111.094434}{Phys. Rev. B \textbf{111}, 094434 (2025)}.

\bibitem{Shangguan2021NP}
  Y. Shangguan, S. Bao, Z.-Y. Dong, N. Xi, Y.-P. Gao, Z. Ma, W. Wang, Z. Qi, S. Zhang, Z. Huang, J. Liao, X. Zhao, B. Zhang, S. Cheng, H. Xu, D. Yu, R. A. Mole, N. Murai, S. O.-Kawamura, L. He, J. Hao, Q.-B. Yan, F. Song, W. Li, S.-L. Yu, J.-X. Li , and J. Wen,
  A one-third magnetization plateau phase as evidence for the Kitaev interaction in a honeycomb-lattice antiferromagnet,
  \href{https://doi.org/10.1038/s41567-023-02212-2}{Nat. Phys. \textbf{19}, 1883–1889 (2023)}.

\bibitem{Khatua2024PRB}
   J. Khatua, M. Gomil\v{s}ek, K.-Y. Choi, and P. Khuntia,
   Magnetism and field-induced effects in the {$S = \frac{5}{2}$} honeycomb lattice antiferromagnet FeP$_3$SiO$_{11}$,
   \href{https://doi.org/10.1103/PhysRevB.110.184402}{Phys. Rev. B \textbf{110}, 184402 (2024)}.

\bibitem{Luo2022PRB}
   Q. Luo, and H.-Y. Kee, 
   Interplay of magnetic field and trigonal distortion in the honeycomb $\Gamma$ model: Occurrence of a spin-flop phase,
   \href{ https://doi.org/10.1103/PhysRevB.105.174435}{Phys. Rev. B \textbf{105}, 174435 (2022)}.

\bibitem{Li2025PRB}
   X. Li, J. Zhao, J. Li, and Q. Luo, 
   Successive topological phase transitions in two distinct spin-flop phases on the honeycomb lattice,
   \href{ https://doi.org/10.1103/PhysRevB.111.064416}{Phys. Rev. B \textbf{111}, 064416 (2025)}.

\bibitem{Xu2020PRL}
   C. Xu, J. Feng, M. Kawamura, Y. Yamaji, Y. Nahas, S. Prokhorenko, Y. Qi, H. Xiang, and L. Bellaiche, 
   Possible Kitaev Quantum Spin Liquid State in 2D Materials with $S = 3/2$,
   \href{https://doi.org/10.1103/PhysRevLett.124.087205}{Phys. Rev. Lett. \textbf{124}, 087205 (2020)}.

\bibitem{Nabi2025arXiv}
   S. D. Nabi, M. Zhu, K. Yu. Povarov, D. G. Mazzone, J. Lass, Y. Wu, Z. Yan, S. Gvasaliya, and A. Zheludev,
   Spin-flop-like transition as quantum critical point in Cs$_2$RuO$_4$,
   \href{https://doi.org/10.48550/arXiv.2507.19853}{arXiv:2507.19853}.

\bibitem{DataAndCode}
   S. Liu, H. Wu, J. Li, X. Wang, and Q. Luo,
   Data repository for ``Emergence of quantum spin liquid and spin-flop phase in Kitaev antiferromagnets in a [111] magnetic field", Zenodo (2025).
    \href{https://doi.org/10.5281/zenodo.17769428}{https://doi.org/10.5281/zenodo.17769428}.

    

\end{thebibliography}
\end{document}